\documentclass{ws-ijmpa}
\usepackage{tikz}
\usetikzlibrary{shapes.geometric, arrows}
\tikzstyle{startstop} = [rectangle, minimum width=1cm, text width=3.5cm, minimum height=1cm,text centered, draw=black]
\tikzstyle{arrow} = [thick,->,>=stealth]
\usepackage{amsmath}
\usepackage{amssymb}

\usepackage[all]{xy}
\begin{document}

\markboth{Y. Dominguez and R. Gaitan} {Elements of the Metric-Affine Gravity I... }

%
\catchline{}{}{}{}{}
%

\title{Elements of the Metric-Affine Gravity I: Aspects of $F(R)$ theories reductions and the Topologically Massive Gravity}

\author{Yessica Dominguez and Rolando Gaitan D.$^{\dag}$}

\address{$^{\dag}$Theoretical and Applied Physics Research Group,\\ A. P. Valencia 2001, Edo. Carabobo,
Venezuela.\\
torsionon@gmail.com}

\maketitle


\begin{abstract}

Some classical aspects of Metric-Affine Gravity  are reviewed in the context of the $F^{(n)}(R)$ type models (polynomials of degree $n$ in the Riemann tensor) and the topologically massive gravity. At the non-perturbative level, we explore the consistency of the field equations when the $ F ^ {(n)} (R) $ models are reduced to a Riemann-Christoffel (RCh) space-time, either
via a Riemann-Cartan (RC) space or via an Einstein-Weyl (EW) space. It is well known for the case $ F^{(1)}(R) = R $ that any
path or reduction ''classes'' via RC or EW, leads to the same field equations with the exception of the $ F^{(n)}(R) $ theories  for $ n> 1 $. We verify that this discrepancy
can be solved by imposing non-metricity and torsion constraints. In particular, we explore the case $ F^{(2)}(R)$ for the interest in expected physical solutions as those of conformally
flat class. On the other hand, the symmetries of the topologically massive gravity are reviewed,
as the physical content in RC and EW scenarios. The appearance of a non-linearly modified selfdual model in RC and existence of many non-unitary degrees of freedom in EW with the suggestion of a modified model for a massive gravity which cure the unphysical propagations, shall be discussed.

\end{abstract}


\section{Introduction}

During the both 60 and 70 decades it emerged the renormalization problems in the Einstein Gravity, which it led to the introduction of Lagrangian counter-terms of higher order, for example the introduction of quadratic terms in the Riemann curvature tensor\cite{Stelle}. This kind of Lagrangians, in the perturbative quantization, it showed Feynman diagrams with less several divergences than the Hilbert-Einstein theory. With those new self-interactions terms via powers of the curvature tensor, other propagations appear besides the graviton, including other physical particles and ghost particles\cite{Barth}. Obviously this particular way of modifying Einstein Theory is part of a bigger problem like the search for consistent formulation of quantum gravity where it is not yet clear from which classical theory of gravity is correct  to start or which the quantum-mechanics framework that we can choose to this task\cite{Gambini} \cite{Gambini2}.

However before of problem statement of the renormalization of the Hilbert-Einstein theory there was intersting in formulation type of $ F^{(2)}(R) $. An important find was, what it was later called the Gauss-Bonnet action, that in $4$ dimensions it is reduced to boundary term\cite{Lanczos} (related with the Euler's index). Later the consistency of certain models $ F^{(2)}(R) $ was studied with the Hilbert-Einstein formulation in terms of their solutions in the vacuum\cite{Stephenson} \cite{Higgs}.

From an experimental-observational point of view the comprobations of General Relativity when fields are weak has dominated. The precession of Mercury's orbit, the existence of gravitational waves observed by the detector of gravitational waves called LIGO (The Laser Interferometer Gravitational-Wave Observatory) and the recent conclusion of the study about the orbit of the star $S2$, with mass 15 times higher than the mass of sun, which has an non-eliptical orbit around of a supermavise black hole in the constellation of Sagittarius. The accelerated expansion of the universe could show that, probably, the gravitation is not described by the Einstein Theory. Instead of this many cosmologists have considered the had hoc way the presence of the dark matter. However, the theories of type $F^{(n)}(R)$ \cite{Sotiriou}, including for $ n < 0 $\cite{Carroll} that it has been already studied from the classical  and (perturbative) quantum theory, today they are candidates to non-perturbative corrections of Einstein gravity with the intention of studing the dark matter problem\cite{Capozziello} \cite{Asgari} \cite{Olmo}.

In addition to this particular interest in modifying the Einstein's theory of gravity through incorporation of  $F^{(n)}(R)$ terms, the extension of the geometry to non-pseudo-Riemannians alternatives  like the Metric-Affine Gravity (MAG) has been take place\cite{Hehl}. Gravity models of higher order with extended geometry, that's means space-time provided with torsion and non-metricity, leads to a bigger number of degrees of freedom. Even when it can bring new non-unitary propagations, those models could lead to a greater quantity of new interaction terms. This new perspective has been using in the extension of the Palatini's variacional principle\cite{Misner} beyond pseudo-Riemannian manifolds\cite{Hehl2} \cite{Ortin}. The extensions of gravity theories not only include new ''cinematic'' parts but some massive terms like in the Topologically Massive Gravity (TMG)\cite{DJT}, for example. Later TMG model was extended to a non-pseudo-Riemannian scenneries\cite{Mielke} \cite{Tresguerres} and it could be called Topologically Massive Metric-Affine Gravity. This model, like the TMG model, are lagrangian models based on a Metric-Affine Hilbert-Einstein action and a term like Chern-Simons. However there is another model\cite{Gaitan1} \cite{Gaitan2} based on a Riemann-Cartan manifold that replaces the $ F^{(1)}(R) $ Lagrangian with a Yang-Mills $ F^{(2)}(R) $ term, whose pseudo-Riemannian limit reproduces the same TMG's field equations.  

This notes are organized as follows. In the next section we perform a brief review on the elements of MAG and notation. We study the role of the affine connection like a Yang-Mills field that transforms under gauge group $ GL(N,R) $  and its natural decomposition in terms of Christoffel's symbols, the contortion tensor and the non-metricity tensor. Section $3$ is devoted to the problem about classical consistency of field equations when a MAG theory type $ F^{(n)}R $ is reduced to a pseudo-Riemannian space (RCh). We verify that the inequivalence that happen when reducing via a Riemann-Cartan space (RC)\cite{Olmo2} or through of an Einstein-Weyl space (EW) it can be avoided by the use of the Lagrange field-multipler technique\cite{Ray} \cite{Safko}. In section $4$ we focus on a general model for a $ F^{(2)}(R) $ theory  with the goal of explore the existence, or not of conformally flat solutions in vacuum, showing that there are forbbiden values parameters within the non-unique definition of $ F^{(2)}(R) $. This sets a classical criterion for a suitable election of some $ F^{(2)}(R) $ term which would be introduced in any theory of gravity. The following sections are devoted to the study of the Physical content of the Topologically Massive Gravity (TMG) in a MAG space. In section $5$ we review the physical content of the well known DJT model in order to contrast its extension to MAG and a brief review on the gauge/diffeomorphism symmetries. Section $6$ is dedicated to explore the physical content of TMG in a RC space confirming the same number of unitary degree of freedom of the Mielke-Baekler model but computting explicitly the total energy and underlining the natural appearance of an extension of the original spin $2$ selfdual model like the original theory\cite{Aragone} \cite{Arias} but with a cubic self-interaction term. The section $7$ is devoted to the study of the TMG in an EW space. Many more degree of freedom appear including tachyonic ghosts among other unphysical manifestations. We briefly discuss the next step to solve these adversities through introduction of a modified massive gravity. Finally we present some concluding remarks.

\section{Coordinates, connections and elements of the Metric-Affine Gravity. Notation.}

It is common to make a distinction between symmetries under change of space-time coordinates and symmetries under gauge changes. But, the latter could be considered as coordinate transformations in some internal space. It is possible to establish a bridge between the idea of space-time coordinates transformation group   and gauge group when it is realized an analogy between group spaces through the concept of connection. In this sense, a first step could be to note the difference between local and non-local transformation. In the context of special relativity, its postulates lead to consider the Lorentz group as a non-local group in the sense that its parameters do not depend of space-time coordinates. Incorporation of non-inertial reference frames opens the window to a new vision of the physical world (General Relativity) where it is possible to establish different kinds of local transformations.

At a point ''$p$'' of a $N$-dimensional space-time manifold, $M^N$, it is possible to distinguish between three classes of transformations which we shall suppose not singular. On the one hand, the Lie group $GL(N,R)$ connects locally curvilinear coordinates $ x^\mu \leftrightarrow {x'}^\mu $ with the prescription
\begin{equation}
{dx'}^\mu = {U^\mu}_\nu (p) dx^\nu \,\,,\,\, \label{ecu:1}
\end{equation}
where ${U^\mu}_\nu (p) = \frac{\partial {x'}^\mu}{\partial x^{\nu}}\mid_{p\in M^N}\,\in\, GL(N,R)$. The coordinates systems connected by $ GL(N,R) $ form an ordered structure of {\it tangent bundle} of the manifold $ M^N $. Next to this coordinates system coexist what we could call {\it Lorentzian} coordinates system, $ \xi ^{a} ; a,b,c = 0,1,\ldots, N-1, $ which transform under local Lorentz transformations, $\mathcal{L}_p\approx ISO(N-1,1)$, or local isometries of the Minkowski's metric, $\eta = diag(-1,1,...,1)$. We mean
\begin{equation}
\eta_{ab} = {(\Lambda^{-1})^c}_a(\xi_p) {(\Lambda^{-1})^d}_b(\xi_p)\eta_{cd}\,\,,\,\, \label{ecu:2}
\end{equation}	
where $ {\Lambda^a}_b (\xi)  \equiv \dfrac{\partial {\xi'}^a}{\partial \xi^b} \arrowvert _{\xi _p} \in \mathcal{L}_p$. This matrix representation of $ \mathcal{L}_p $ allows us to write the change of local coordinates in $p$
\begin{equation}
{\xi'}^a |_{\xi_p} = {\Lambda^a}_b (\xi_p) \xi^b \,\,.\,\,\label{ecu:3}
\end{equation}

A third kind of transformation what we want to highlight connect curvilinear coordinates $ \xi^\mu $ with local Lorentzian coordinates , $ \xi^a$, for which we introduce the {\it vielbeins}, $ {\mathrm{e}^a}_\mu $ through
\begin{equation}
dx^\mu |_{x = x(\xi_p)} =  {\mathrm{e}_a}^\mu (\xi_p)d\xi^a \,\,,\,\, \label{ecu:4}
\end{equation}	
where $ {\mathrm{e}_a}^\mu (\xi_p) \equiv \dfrac{\partial x^\mu}{\partial \xi^a}|_{\xi_p} $ y $ {\mathrm{e}^a}_\mu (\xi_p) \equiv \dfrac{\partial \xi^a}{\partial x^\mu}|_{\xi_p} $. The vielbeins are useful to project a general metric over a local Lorentzian frame of the form
\begin{equation}
\eta _{ab} = {\mathrm{e}_a}^\mu (p) {\mathrm{e}_b}^\nu (p) g_{\mu\nu}(x_p) \,\,.\,\,\label{ecu:5}
\end{equation}

The transformation groups mentioned in \eqref{ecu:1}, \eqref{ecu:3} and \eqref{ecu:4} can be schematized by the identity between maps, as 
\begin{equation}
\mathrm{e}^{-1} \Lambda \mathrm{e}U^{-1} = \mathbb{I}\,\,,\,\,\label{ecu:5a}
\end{equation}
\begin{equation}
\xymatrix{
x^\mu \ar[d]_U & \ar[l]_{\mathrm{e}} \xi^a \ar[d]^\Lambda \\
{x'}^\mu & {\xi'}^a \ar[l]^{\mathrm{e}}
} \nonumber
\end{equation}
from where, it get the following equivalence of compositions, $ \mathrm{e}^{-1}\Lambda = U\mathrm{e}^{-1} $, indicating, for example that a point $ p $ of $ M $ with coordinates $ {{\xi'}^a}_p $ can be described locally with general curvilinear coordinates, i.e. $ {\xi'}^a{_p} = {\xi'}^a({x'}^\mu(x^\mu)) = {\xi'}^a(\xi^a(x^\mu)) $.

An operational way to introduce the idea of connection is to take the group of local Lorentz transformations, $ \mathcal{L}_p $ whose elements act as $ {t'}_a (\xi)|_{\xi'(\xi_p)} = {(\Lambda)^b}_a (\xi_p)t_b(\xi_p) $, therefore (thanks to identity \eqref{ecu:5a}) we can attach this one to local curvilinear coordinates  
\begin{equation}
{T'}^a (x')|_{x'(x)} = {(\Lambda^{-1})^b}_a(\xi(x)) T_b(x)\,\,,\,\,\label{ecu:6}
\end{equation}
Where $ T^a (x)\equiv t^a (\xi(x)) $ y $ {T'}^a (x')|_{x'(\xi)} \equiv {t'}^a(\xi'(x)) $. Next, we introduce a covariant derivative under $ \mathcal{L}_p $ as follows
\begin{equation}
D^{(\omega)}_\mu  T_a = \partial_\mu T_a - {\omega^b}_{\mu a} T_b\,\,,\,\, \label{ecu:7}
\end{equation}
where ${\omega^b}_{\mu a}\equiv {(\omega_\mu)^b}_a $ is the matrix component of the {\it Lorentz connection}, $ \omega_\mu $. We obtain the transformation rule of the Lorentz connection, under $ \mathcal{L}_p $, imposing the covariance of the equation \eqref{ecu:7}, this means $ D^{(\omega)}_\mu  T_a)' = {(U^{-1})^\alpha}_\mu {(\Lambda^{-1})^b}_a D^{(\omega)}_\alpha  T_b $ from which we write  
\begin{equation}
{{\omega'}^a}_{\mu b} = {(U^{-1})^\alpha}_\mu \big[{\Lambda^a}_d{\omega^d}_{\alpha c}{(\Lambda^{-1})^c}_b + {\Lambda^a}_d\partial_\alpha {(\Lambda^{-1})^d}_b\big]\,\,.\,\, \label{ecu:8}
\end{equation}		
If we define $ {({\omega'}_\beta)^a}_b \equiv {U^\mu}_\beta {{\omega'}^a}_{\mu b} $ as the Lorentz connection with $ \beta $ projected in the coordinates system $ x^\mu $, we can rewrite the equation \eqref{ecu:8} in matrix representation as
\begin{equation}
\omega'_\mu = \Lambda \omega_\mu\Lambda^{-1} + \Lambda \partial _\mu \Lambda^{-1}\,\,,\,\,\label{ecu:9}
\end{equation}
and in this way we will look at the group, $ \mathcal{L} _p $ , as a gauge group and $\omega_\mu$ is the Yang-Mills field associated to the Lorentz symmetry.

Just as it is possible to build gravity theories as gauge theories based on local Lorentz symmetry (as in the original topologically massive gravity version\cite{DJT}) taking Lorentz connection and correspondent curvature, $ R_{\mu \nu} = \partial_\mu \omega_\nu - \partial_\nu \omega_\mu + [\omega_\mu, \omega_\nu] $, our interest is to extend this idea by considering the group $GL(N,R)$ of general transformations as a gauge group. As we have previously established, a $N$-dimensional manifold allows, among others things an additional structure as the affine-connection idea whose matrix representation is $ {(A_\mu)^\alpha}_\beta \equiv {A^\alpha}_{\mu\beta} $. As in the same way that occurs with the Lorentz symmetry, here we introduce the covariant derivative 
\begin{equation}
D^{(A)}_\mu t_\nu = \partial_\mu t_\nu - {A^\lambda}_{\mu \nu} t_\lambda \,\,,\,\,\label{ecu:10}
\end{equation}
which, being covariant under  $ GL(N,R) $ leads to the following transformation rule for affine-connection 
\begin{equation}
A'_\mu = UA_\mu U^{-1} + U\partial_\mu U^{-1} \,\,,\,\,\label{ecu:11}
\end{equation} 
where $ U \in GL(N,R)$. In the notation, $ A'_\mu  $ means that the index $ \mu $ lies in the old coordinates system $ x^\mu $, meanwhile matrix representation indices are in the new coordinate system $ {x '}^\mu $. 

Without making any kind of restriction on the geometry let us to recall some MAG elements. The torsion tensor can be defined as 
\begin{equation}
{T^\lambda}_{\mu \nu} \equiv {A^\lambda}_{\mu \nu} - {A^\lambda}_{\nu \mu} = - {T^\lambda}_{\nu\mu}\,\,,\,\,  \label{ecu:13}
\end{equation}
whose tensor character is guaranteed since it comes from the difference of connections on the same fiber. The other relevant object is known as the tensor of non-metricity tensor, which is defined as the covariant derivative of the metric tensor, that is
\begin{equation}
Q_{\alpha \mu \nu}\equiv D_\alpha g_{\mu \nu} = Q_{\alpha\nu\mu}\,\,.\,\, \label{ecu:14}
\end{equation} 

With the definition of torsion and non-metricity given by the equations \eqref{ecu:13} and \eqref{ecu:14} it is possible to describe different kinds of space-time and its reductions. In this sense, the definition of non-metricity \eqref{ecu:14} allow us to establish the following decomposition for the affine-connection
\begin{equation}
{A^\lambda}_{\mu\nu} = {\Gamma^\lambda}_{\mu\nu}(g) + {K^\lambda}_{\mu \nu}(T) - {\gamma ^\lambda}_{\mu \nu}(Q)\,\,,\,\,\label{ecu:15}
\end{equation}
where
\begin{equation}
{\Gamma^\lambda}_{\mu\nu}(g) \equiv \dfrac{g^{\lambda \sigma}}{2}\,(\partial_\mu g_{\sigma \nu} + \partial_\nu g_{\sigma\mu} - \partial_\sigma g_{\mu \nu})\,\,,\,\,\label{ecu:16}
\end{equation}
are the components of the Levi-Civita connection or Christofell symbols. The next object
\begin{equation}
{K^\lambda}_{\mu \nu}(T) \equiv \dfrac{1}{2}\,({{T_\mu}^{\lambda}}\,_\nu + {{T_\nu}^{\lambda}}_\mu + {T^\lambda}_{\mu\nu} )\,\,,\,\, \label{ecu:17}
\end{equation}
are the components of the contortion tensor with ${K^\lambda}_{\mu \nu}=-{K_{\nu\mu}}^\lambda$ and the last ones
\begin{equation}
{\gamma^\lambda}_{\mu \nu}(Q) \equiv \dfrac{g^{\lambda \alpha}}{2}\,(Q_{\mu\alpha\nu} + Q_{\nu\alpha\mu} - Q_{\alpha\mu\nu} ) \,\,,\,\,\label{ecu:18}
\end{equation}
we shall call them {\it non-metricity symbols}. The concepts of torsion and non-metricity allow to classify different geometries (Fig.1) on which we will make observations throughout this work.
\begin{figure}
\begin{tikzpicture}[node distance=1cm]
\node (start)[startstop]{Metric-Affine Gravity};
\node (start1a)[startstop, below of=start, yshift=-2cm, xshift=4cm]{Einstein-Weyl};
\node (start1b)[startstop, below of=start, yshift=-2cm, xshift=-4cm]{Riemann-Cartan};
\node (start2)[startstop, below of=start1a, yshift=-2cm, xshift=-4cm]{Riemann-Christoffel};
\draw [arrow] (start) -- node[anchor=west] {${\bf T=0}$} (start1a);
\draw [arrow] (start) -- node[anchor=east] {${\bf Q=0}$} (start1b);
\draw [arrow] (start1a) -- node[anchor=west] {${\bf Q=0}$} (start2);
\draw [arrow] (start1b) -- node[anchor=east] {${\bf T=0}$} (start2);
\draw [arrow] (start) -- node[anchor=west] {} (start2);
\end{tikzpicture}
\caption{}
\end{figure}

The next step consits on review the habitual action principle in a metric-affine space, particularly in the case of vacuum. From the point of view  of the Lagrangian field theory we considere a configuration space with field-coordinates and field-velocities provided, $ \{g_{\mu \nu}; \partial_\alpha g_{\mu \nu};{A^\lambda}_{\mu \nu}; \partial_\alpha {A^\lambda}_{\mu \nu}\}$ which allow us to address first-order formulations like a Lagrangian density proportional to the scalar curvature $R$. Therefore, we would say that a motivation for this approach is the direct identification of the conjugate canonical variables taking in sight a possible quantum-mechanical study.

Thus, a starting non-massive and source free model shall be the Lagrangian density given by the scalar $ R(a,A) = g^{\mu\nu} R_{\mu\nu} = g^{\mu \nu} R^{\lambda}{_{\mu \lambda \nu}}(A) $, where the Riemann-metric-affine curvature has a standard matrix form like $\mathbf{R}_{\sigma\nu} = \partial_\nu A_\sigma - \partial_\sigma A_\nu + [A_\nu, A_\sigma]$ or in components $
{R^\lambda}_{\mu\sigma\nu} (A) = {(\mathbf{R}_{\sigma\nu})^{\lambda}}_\mu = \partial_\nu {A^\lambda}_{\sigma\mu} - \partial_\sigma {A^\lambda}_{\nu\mu} + {A^\lambda}_{\nu\rho}{A^\rho}_{\sigma\mu} - {A^\lambda}_{\sigma\rho}{A^\rho}_{\nu\mu}$. Using \eqref{ecu:15} we can rewrite the Riemann curvature as follows
\begin{equation}
{\mathbf{R}}_{\sigma\nu}= {\mathbf{R}}_{\sigma\nu}(\Gamma) + {\mathbf{R}}_{\sigma\nu}(\tau) + 2[\Gamma_{[\nu }, \tau_{\sigma]}]\,\,.\,\, \label{ecu:22}
\end{equation}
where we can identify the Riemann-Christoffel curvature
\begin{equation}
{\mathbf{R}}_{\sigma\nu}(\Gamma) \equiv \partial_\nu \Gamma_\sigma - \partial _\sigma\Gamma_\nu + [\Gamma_\nu, \Gamma_\sigma]\,\,,\,\,\label{ecu:21}
\end{equation}
and 
\begin{equation}
\tau _\mu \equiv K_\mu - \gamma_\mu\,\,,\,\,\label{ecu:21a}
\end{equation}
as the {\it geodesic deviation tensor}.

The first contraction  of \eqref{ecu:22}, this means ${R^\mu}_{\lambda\mu\nu}$ provides the Ricci-metric-affine tensor
\begin{equation}
R_{\lambda\nu} = R_{\lambda\nu}(\Gamma) + {[\tau_\nu,\tau_\sigma]^\sigma}_\lambda + \nabla_\nu {\tau^\sigma}_{\sigma\lambda} - \nabla_\sigma {\tau^\sigma}_{\nu\lambda}\,\,,\,\, \label{ecu:24}
\end{equation}
and contracting again, the Ricci-metric-affine scalar
\begin{equation}
R = R(\Gamma) + g^{\nu\lambda}{[\tau_\nu,\tau_\sigma]^\sigma}_\lambda + \nabla_\nu {{\tau^\sigma}_\sigma}^\nu - \nabla_\sigma {{\tau^\sigma}_\nu}^\nu \,\,.\,\, \label{ecu:25}
\end{equation}
Now, we immediately think in the vacuum-free action that it would have by name Hilbert-Einstein-metric-affine theory (up to a boundary term)
\begin{equation}
S_0 = S_{HE} -k^{2-N} \langle  g^{\nu \lambda} {[\tau_\nu, \tau_\sigma]^\sigma}_\lambda \rangle \,\,,\,\,\label{ecu:26}
\end{equation}
where $ k $ has length units.  It can be observed that the geodesic deviation will participate explicitly in the coupling between gravitons, torsion and non-metricity.

\section{Classical consistency of metric-affine $F^{(n)}(R)$ models.}

Here we are interested in checking the consistency of the field equations which are derived from a Lagrangian of the form $F^{(n)}(R)$ when the metric-affine manifold is "reduced" according to any path in the scheme of figure 1. This problem has already been reported in the literature and our main purpose is to systematize these observations. The other aspect that we wish to underline is the existence or not of conformally planar solutions according to the particular characteristics of the Lagrangian in the $n=2$ case. 

\subsection{Aspects of the reduction of the metric-affine manifold for $n=1$.}

In many references \cite{Ray} \cite{Safko} \cite{Olmo} related to the gravity's variational principle, there has been a lot of discussion about the effect that torsion has at the level of field equations when these are contrasted with those obtained after the application of the Palatini method, which originally supposes a priori a symmetric connection. Beyond the Hilbert-Einstein Lagrangian formulation and considering theories of the class $F^{(n)}(R)$ with $n>1$, the discrepancy of the equations of motion has been reported when the condition of null torsion is imposed before or after performing arbitrary functional variations on fields.

With this motivation, we will review the study on the classification of the various MAG's reduction pathways towards a Riemann-Christoffel (RCh) manifold. Discussion around the Palatini principle and its variant with null torsion has been carried out for a long time, which corresponds to a reduction path of the EW class (Einstein-Weyl: see figure 1) including the Lagrange multipliers technique\cite{Ray} \cite{Safko}. 

First at all we want to review the  $F^{(1)}(R)$ formulation for source free action in order to establish the notation of the variational principles that we will implement according to the reduction path classes shown in figure 1.

The MAG$\rightarrow$EW class of reduction paths contains three ways which can be distinguished: (i) imposing the torsionless condition before performing the variational principle (Palatini), (ii) taking the null torsion in the MAG field equations or (iii) including a Lagrangian constraint before of making variations, verifying that for $F^{(1)}(R)$ case, any  path lead to null non-metricity. Next we will introduce the torsion constraint through the action
\begin{eqnarray}
S'^{(1)}=k^{2-N}\, \langle g^{\mu\nu}R_{\mu\nu}+
{b_\alpha}^{\mu\nu}{T^\alpha}_{\mu\nu}\,\rangle
\,\, ,\,\, \label{3-1}
\end{eqnarray}
where ${b_\alpha}^{\mu\nu}=-{b_\alpha}^{\nu\mu}$ is a Lagrangian multiplier field. Arbitrary variations on (\ref{3-1}) mean $\delta S'^{(1)}=k^{2-N}\, \langle {\tau^{(1)}}_{\mu\nu} \delta g^{\mu\nu}-2({{E^{(1)}}_\alpha}^{\mu\beta}-{b_\alpha}^{\mu\beta})\delta {A^\alpha}_{\mu\beta} \rangle$ and give rise the following field equations
\begin{eqnarray}
{\tau^{(1)}}_{\mu\nu}=0  \,\, ,\,\, \label{3-2}
\end{eqnarray}
\begin{eqnarray}
{{E^{(1)}}_\alpha}^{\mu\beta}={b_\alpha}^{\mu\beta} \,\, ,\,\, \label{3-3}
\end{eqnarray}
\begin{eqnarray}
{T^\alpha}_{\mu\nu}=0 \,\, ,\,\, \label{3-4}
\end{eqnarray}
where we have defined
\begin{eqnarray}
{\tau^{(1)}}_{\mu\nu}\equiv R_{\{\mu\nu\}}-\frac{g_{\mu\nu}}{2}\,R  \,\, ,\,\, \label{3-5}
\end{eqnarray}
\begin{eqnarray}
{{E^{(1)}}_\alpha}^{\nu\lambda}\equiv -\frac{1}{2} \Big({Q_\sigma}^{\lambda\nu}-{\delta^\nu}_\sigma {Q_\beta}^{\lambda\beta}
+\frac{g_{\alpha\beta}}{2}\,(g^{\lambda\rho}{\delta^\nu}_\sigma - g^{\lambda\nu}{\delta^\rho}_\sigma){Q_\rho}^{\alpha\beta} +\nonumber \\
+\,{\delta^\nu}_\sigma{{T^\beta}_\beta}^\lambda- 
g^{\lambda\nu}{T^\beta}_{\beta\sigma}+{T^{\nu\lambda}}_\sigma \Big) \,\, .\,\, \label{3-6}
\end{eqnarray}

The unique solution for \eqref{3-2}, \eqref{3-3} and  \eqref{3-4} is
\begin{eqnarray}
R_{\mu\nu}-\frac{g_{\mu\nu}}{2}\,R =0  \,\, ,\,\, \label{3-7}
\end{eqnarray}
\begin{eqnarray}
{T^\alpha}_{\mu\nu}=0 \,\, ,\,\, \label{3-8}
\end{eqnarray}
\begin{eqnarray}
Q_{\alpha\mu\nu}=0 \,\, ,\,\, \label{3-9}
\end{eqnarray}
\begin{eqnarray}
{b_\alpha}^{\mu\beta}=0 \,\, ,\,\, \label{3-10}
\end{eqnarray}
confirming the reduction to a Riemann-Christoffel space-time without the need neither to introduce the torsionless constraint, nor to impose the null torsion condition strongly, before or after the realization of the variational principle. Furthermore, if we inspect the first functional variation of the unconstained action in the metric-affine space, in other words $\delta S^{(1)}=k^{2-N}\, \langle {\tau^{(1)}}_{\mu\nu} \delta g^{\mu\nu}-2{{E^{(1)}}_\alpha}^{\mu\beta}\delta {A^\alpha}_{\mu\beta} \rangle$ and later setting  ${T^\alpha}_{\mu\nu}=0$ on MAG field equations, it can be obtained again \eqref{3-7} and \eqref{3-9}. If instead we assume in advance a null torsion, the first functional variation of the action is $\delta S^{(1)}=k^{2-N}\, \langle {\tau^{(1)}}_{\mu\nu}\mid_{T=0}\, \delta g^{\mu\nu}-2{{E^{(1)}}_\alpha}^{\{\mu\beta\}}\mid_{T=0}\,\delta {A^\alpha}_{\{\mu\beta\}} \rangle$, being able to show that the ''new'' equations of motion lead again to both \eqref{3-7} and \eqref{3-9}.

Now we want to explore the MAG$\rightarrow$RC class of reduction paths  (fig. 1). Let us start with the reduction path which take in acount null non-metricity before performing functional variations. This means to impose the next $\frac{N^2(N+1)}2$ constraints
\begin{eqnarray}
Q_{\alpha\mu\nu} =\partial_\alpha g_{\mu\nu}-{A^\lambda}_{\alpha\mu}\,g_{\lambda\nu}-{A^\lambda}_{\alpha\nu}\,g_{\lambda\mu}=0
\,\, ,\,\, \label{3-11}
\end{eqnarray}
and from this relation it can be shown that symmetric part of affine connection depends on metric and torsion as follows
\begin{eqnarray}
{A^\nu}_{\{\alpha\mu\}}={\Gamma^\nu}_{\alpha\mu}+\frac{g^{\nu\lambda}g_{\mu\sigma}}2\,{T^\sigma}_{\lambda\alpha}+\frac{g^{\nu\lambda}g_{\alpha\sigma}}2\,{T^\sigma}_{\lambda\mu}
\,\, ,\,\, \label{3-12}
\end{eqnarray}
it will be helpful to know its functional variation
\begin{eqnarray}
\delta {A^\nu}_{\{\alpha\mu\}}=\Gamma_{\rho\alpha\mu} \delta g^{\nu\rho} +\frac{g^{\nu\rho}}2\,(\partial_\alpha \delta g_{\rho\mu}
+\partial_\mu \delta g_{\rho\alpha}-\partial_\rho \delta g_{\alpha\mu})
+T_{\{\mu\lambda\alpha\}}\delta g^{\nu\lambda}+\nonumber \\
+ 
{T^{\sigma\nu}}_{\{\alpha} \delta g_{\sigma\mu\}}+g^{\nu\lambda}g_{\mu\sigma}\delta {A^\sigma}_{[\lambda\alpha]}+g^{\nu\lambda}g_{\alpha\sigma}\delta {A^\sigma}_{[\lambda\mu]}
\,\, .\,\, \label{3-13}
\end{eqnarray}

Thus with the help of \eqref{3-13} the first functional variation of $S^{(1)}\mid_{Q=0}=k^{2-N}\, \langle g^{\mu\nu}R_{\mu\nu}\rangle \mid_{Q=0}$ is written as 
\begin{eqnarray}
\delta \big( S^{(1)}\mid_{Q=0}\big)=k^{2-N}\, \langle \big({\tau^{(1)}}_{\mu\nu}-D_\alpha {P^{(1)\alpha}}_{\mu\nu}+{T^\alpha}_{\alpha\sigma} {P^{(1)\sigma}}_{\mu\nu}\big)\delta g^{\mu\nu}+\nonumber \\
-2\big({{E^{(1)}}^{[\alpha\beta]}}_{\sigma}+{{{E^{(1)}}^{[\alpha}}\,_\sigma}\,\,^{\beta]}+
{{E^{(1)}}_\sigma}^{[\alpha\beta]}\big)\delta {A^\sigma}_{[\alpha\beta]}\rangle\mid_{Q=0}
\,\, ,\,\, \label{3-15}
\end{eqnarray}
where we have defined
\begin{eqnarray}
P^{(1)\alpha\mu\nu}\equiv 2{E^{(1)}}^{\mu\{\alpha\nu\}}-{E^{(1)}}^{\alpha\{\mu\nu\}}
\,\, ,\,\, \label{3-16}
\end{eqnarray}
and ${E^{(1)}}^{\rho\alpha\beta}$ is given at \eqref{3-6} and some manipulation on field equations which arise from \eqref{3-15} provides the expected result
\begin{eqnarray}
{\tau^{(1)}}_{\mu\nu}=0 \,\, ,\,\, \label{3-17}
\end{eqnarray}
\begin{eqnarray}
{T^\alpha}_{\mu\nu}=0 \,\, .\,\, \label{3-18}
\end{eqnarray}

In contrast to the above, the Lagrange multiplier technique demands the introduction of a non-metricity constraint through multiplier field $a^{\alpha\mu\nu}=a^{\alpha\nu\mu}$ as we show next
\begin{eqnarray}
S''^{(1)}=k^{2-N}\, \langle g^{\mu\nu}R_{\mu\nu}+
a^{\alpha\mu\nu}Q_{\alpha\mu\nu}\,\rangle
\,\, .\,\, \label{3-19}
\end{eqnarray}
The exercise of obtaining the field equations provides ${\tau^{(1)}}_{\mu\nu}={T^\alpha}_{\mu\nu}=Q_{\alpha\mu\nu}=a^{\alpha\mu\nu}=0$.

Finishing this section we want to point out some remarks. On the  one hand it can also be shown that the direct reduction path MAG$\rightarrow$RCh would be performed by means of the action $S'''^{(1)}=k^{2-N}\, \langle g^{\mu\nu}R_{\mu\nu}+
a^{\alpha\mu\nu}Q_{\alpha\mu\nu} + {b_\alpha}^{\mu\nu}{T^\alpha}_{\mu\nu}\,\rangle$ which (nothing surprising) gives the same results again: Einstein's equations are obtained for a pseudo-Riemannian manifold and all multipliers are zero. It seems too trivial and repetitive what has been stated here, fundamentally due to the fact that it does not seem to matter what one does, the Lagrange multipliers are always zero, which indicates that deep down they were never needed. 

It can be shown immediately that the set of multipliers used here, that is $a^{\alpha \mu \nu}$ and ${b_\alpha}^{\mu \nu}$ is equivalent to the set of multipliers, ${\lambda_\alpha}^{\mu\nu}\neq {\lambda_\alpha}^{\nu\mu}$ when the constraint ${A^\alpha}_{\mu\nu}-{\Gamma^\alpha}_{\mu\nu}$ is used\cite{Ray}.
In fact, the multipliers of non-metricity and torsion constraints are related to those of J. R. Ray\cite{Ray} through
\begin{eqnarray}
a^{\alpha\mu\nu}=\frac{1}2\,\big(\lambda^{\alpha\{\mu\nu\}}-\lambda^{\{\mu\alpha\nu\}}- \lambda^{\{\mu\nu\}\alpha}  \big) \,\, ,\,\, \label{3-19a}
\end{eqnarray}
\begin{eqnarray}
b^{\alpha\mu\nu}=\frac{1}2\,\big(\lambda^{\alpha[\mu\nu]}-\lambda^{[\mu\alpha\nu]}- \lambda^{[\mu\nu]\alpha}  \big) \,\, .\,\, \label{3-19b}
\end{eqnarray}

Finally, because the Lagrangian density of the action $S^{(1)}$ is proportional to the Hilbert-Einstein MAG one, it does not matter what kind of reduction paths from MAG we use. However, there are physical systems where the Lagrange multipliers are not trivial\cite{Gaitan2} due to the presence of quadratic terms in the curvature and of massive type in the action (which is not the case we have looked at), which goes hand in hand with the fact that the field equations depend on which reduction path we choose. In any case, the important thing here is that we have established the different restrictions on the variational principle that we will implement in the next section where we shall observe that in effect the lagrange multipliers are no longer trivial.

\subsection{Consistency of field equations in the $F^{(n)}(R)$ model.}

Here we will study the reduction of the metric-affine manifold in the context of a $ F^{(n)}(R)$ Lagrangian density  with $n>1$. Let $ S^{(n)}$ be the action 
\begin{eqnarray}
S^{(n)} =k^{2n-N} \,\langle F^{(n)}(g,R) \rangle
\,\, ,\,\, \label{3-20}
\end{eqnarray}
and its functional variation is
\begin{equation}
\delta S^{(n)} =k^{2n-N} \, \langle \tau_{\mu\nu}\delta g^{\mu\nu}+{F_\alpha}^{\beta\mu\nu}\,\delta {R^\alpha}_{\beta\mu\nu}\rangle \,\, ,\,\, \label{3-21}
\end{equation}
where
\begin{eqnarray}
\tau_{\mu\nu}\equiv \frac{\partial F^{(n)}}{\partial
g^{\mu\nu}}-\frac{F^{(n)}}{2}\,g_{\mu\nu}  \,\, ,\,\, \label{3-22}
\end{eqnarray}
\begin{eqnarray}
{F_\alpha}^{\beta\mu\nu}\equiv \frac{\partial F^{(n)}(R)}{\partial {R^\alpha}_{\beta\mu\nu}}\,\, .\,\, \label{3-23}
\end{eqnarray}

However, remembering the definition of curvature one arrive to $\delta {R^\alpha}_{\beta\mu\nu}=D_\nu \delta {A^\alpha}_{\mu\beta}-D_\mu \delta {A^\alpha}_{\nu\beta}+{T^\sigma}_{\nu\mu}\delta {A^\alpha}_{\sigma\beta}$ and then \eqref{3-21} can be written  in explicit terms of connection variations as follows 
\begin{eqnarray}
\delta S^{(n)} =k^{2n-N} \, \langle \tau_{\mu\nu}\delta g^{\mu\nu}
-2{E_\alpha}^{\sigma\beta}\delta{A^\alpha}_{\sigma\beta}\rangle 
\,\, ,\,\, \label{3-25}
\end{eqnarray}
where the Palatini's tensor is
\begin{eqnarray}
{E_\alpha}^{\sigma\beta}\equiv D_\nu
{F_\alpha}^{\beta\sigma\nu}+\big( \frac{{\delta^\sigma}_\mu g^{\lambda\rho}}{2}\,
Q_{\nu\lambda\rho}-{\delta^\sigma}_\mu{T^\lambda}_{\lambda\nu}+\frac{1}{2}\,{T^\sigma}_{\mu\nu}\big){F_\alpha}^{\beta\mu\nu}
 \,\, .\,\, \label{3-26}
\end{eqnarray}

From this notation we will be able to look at various scenarios and thus contrast the three classes of MAG reduction paths shown in fig.1.

\subsubsection{Einstein-Weyl Class (EW)}

This is the Palatini's way where we assume the torsionless condition before making functional variations. So, from \eqref{3-25} we get
\begin{eqnarray}
\delta \big(S^{(n)}\mid_{T=0}\big) =k^{2n-N} \, \langle \tau_{\mu\nu}\delta g^{\mu\nu}
-2{E_\alpha}^{\{\sigma\beta\}}\delta{A^\alpha}_{\{\sigma\beta\}}\rangle 
\,\, ,\,\, \label{3-29}
\end{eqnarray}
from where the field equations are written (using \eqref{3-26})
\begin{eqnarray}
\tau_{\mu\nu}=0
\,\, ,\,\, \label{3-30}
\end{eqnarray}
\begin{eqnarray}
{E_\alpha}^{\{\sigma\beta\}}\equiv D^{(T=0)}_\nu
{F_\alpha}^{\{\beta\sigma\}\nu} +\frac{g^{\lambda\rho}}{2}\,
Q_{\nu\lambda\rho}\,{F_\alpha}^{\{\beta\sigma\}\nu}=0
 \,\, ,\,\, \label{3-31}
\end{eqnarray}
and the ad hoc condition ${T^\sigma}_{\mu\nu}=0$. On the other hand, if we evaluate the metric-affine equation coming from \eqref{3-25} over the torsionless condition, we obtain 
\begin{eqnarray}
{E_\alpha}^{\sigma\beta}\equiv D^{(T=0)}_\nu
{F_\alpha}^{\beta\sigma\nu}+\frac{g^{\lambda\rho}}{2}\,
Q_{\nu\lambda\rho}\,{F_\alpha}^{\beta\sigma\nu}=0
 \,\, ,\,\, \label{3-32}
\end{eqnarray}
which is algebraically different with respect to \eqref{3-31}. This is a characteristic reported in the literature\cite{Olmo2}. Here we underline for an arbitrary Lagrangian density $ F^{(n)}(R) $, none of the equations \eqref{3-31} or \eqref{3-32} guarantees that the non-metricity is necessarily zero. Only the case $ n=1$ guarantees that the torsionless condition implies the non-metricity is also null.

Nevertheless there exists a third reduction path in the EW class which consists of incorporating the torsionless constraint through the action
\begin{eqnarray}
S'=S^{(n)}+k^{2n-N}\, \langle {b_\alpha}^{\mu\nu}{T^\alpha}_{\mu\nu}\,\rangle
\,\, ,\,\, \label{3_33}
\end{eqnarray}
where ${b_\lambda}^{\mu\nu}=-{b_\lambda}^{\nu\mu}$. The field equations are
\begin{eqnarray}
\tau_{\mu\nu}=0
\,\, ,\,\, \label{3-34}
\end{eqnarray}
\begin{eqnarray}
{E_\alpha}^{\{\sigma\beta\}}\equiv D^{(T=0)}_\nu
{F_\alpha}^{\{\beta\sigma\}\nu}+\frac{g^{\lambda\rho}}{2}\,
Q_{\nu\lambda\rho}\,{F_\alpha}^{\{\beta\sigma\}\nu}=0
 \,\, ,\,\, \label{3-35}
\end{eqnarray}
\begin{eqnarray}
{T^\lambda}_{\mu\nu}=0
 \,\, ,\,\, \label{3-36}
\end{eqnarray}
\begin{eqnarray}
{b_\lambda}^{\mu\nu}={E_\lambda}^{[\mu\nu]}
 \,\, .\,\, \label{3-37}
\end{eqnarray}

It can be noted that \eqref{3-34} and \eqref{3-35} match with the Palatini's method and Lagrange multipliers are no longer necessarily trivial for $n>1$. However, no path of the EW class leads to the equations expected in the pseudo-Riemannian manifold, because even if the limit $Q _{\alpha \mu \nu} \rightarrow 0$ were taken on the equations \eqref{3-31} and \eqref{3-32} we would get $\nabla_\nu {F_\alpha}^{\{\beta\sigma\}\nu} = 0$ and $\nabla_\nu{F_\alpha}^{\beta\sigma\nu} = 0 $, respectively.

\subsubsection{Riemann-Cartan Class (RC)}

Now we explore the reduction paths which supose the $Q_{\alpha\mu\nu}=0$ constraint. The last relation was stablished at \eqref{3-11} and induces a constraint on symmetric part of connection (see \eqref{3-12}), this means ${A^\nu}_{\{\alpha\mu\}}={\Gamma^\nu}_{\alpha\mu}+\frac{g^{\nu\lambda}g_{\mu\sigma}}2\,{T^\sigma}_{\lambda\alpha}+\frac{g^{\nu\lambda}g_{\alpha\sigma}}2\,{T^\sigma}_{\lambda\mu}$. Copying us from \eqref{3-15}, the next functional variation can be written
\begin{eqnarray}
\delta \big( S^{(n)}\mid_{Q=0}\big)
&= k^{2n-N}\, \langle \big(\tau_{\mu\nu}-D_\alpha {P^\alpha}_{\mu\nu}+{T^\alpha}_{\alpha\sigma} {P^\sigma}_{\mu\nu}\big)\delta g^{\mu\nu}+\nonumber \\
&-2\big({E^{[\alpha\beta]}}_{\sigma}+{{E^{[\alpha}}_\sigma}\,^{\beta]}+
{E_\sigma}^{[\alpha\beta]}\big)\delta {A^\sigma}_{[\alpha\beta]}\rangle
\,\, ,\,\, \label{3-38}
\end{eqnarray}
where
\begin{eqnarray}
P^{\alpha\rho\beta}\equiv 2E^{\rho\{\alpha\beta\}}-E^{\alpha\{\rho\beta\}}
\,\, ,\,\, \label{3-39}
\end{eqnarray}
with $E^{\rho\alpha\beta}$ given at \eqref{3-26}.

From \eqref{3-38}, the torsion field equation is 
\begin{eqnarray}
{E^{[\alpha\beta]}}_{\sigma}+{{E^{[\alpha}}_\sigma}\,^{\beta]}+{E_\sigma}^{[\alpha\beta]}=0
\,\, ,\,\, \label{3-40}
\end{eqnarray}
which is equivalent to 
\begin{eqnarray}
E_{[\alpha\sigma\beta]}=0
\,\, ,\,\, \label{3-41}
\end{eqnarray}
or equal to $E_{\alpha\{\sigma\beta\}}-E_{\{\sigma\beta\}\alpha}=0$ and this allow us to rewrite \eqref{3-39} as follows 
\begin{eqnarray}
P^{\alpha\rho\beta}= E^{\rho\alpha\beta}-E^{\alpha[\rho\beta]}
\,\, ,\,\, \label{3-42}
\end{eqnarray}
hence metric and connection field equations are 
\begin{eqnarray}
\tau_{\mu\nu}-D_\alpha {{E_\mu}^\alpha}\,_\nu+{T^\alpha}_{\alpha\sigma} {{E_\mu}^\sigma}\,_\nu=0
\,\, ,\,\, \label{3-43}
\end{eqnarray}
\begin{eqnarray}
E_{[\alpha\sigma\beta]}\equiv D_\nu {F_{[\alpha\beta]\sigma}}^\nu+\big(-g_{\sigma\mu}\,{T^\lambda}_{\lambda\nu}+\frac{1}{2}\,T_{\sigma\mu\nu}\big){F_{[\alpha\beta]}}^{\mu\nu}=0
 \,\, ,\,\, \label{3-44}
\end{eqnarray}
and the ad hoc constraint
\begin{eqnarray}
Q_{\nu\lambda\rho}=0
 \,\, .\,\, \label{3-45}
\end{eqnarray}
It can be noted that the torsionless limit of \eqref{3-43} does not have to match with the metric field equation in a  pseudo-Riemannian manifold. On the other hand, considering ${T^\lambda}_{\mu\nu}=0$ in \eqref{3-44} leads to $\nabla_\nu {F_{[\alpha\beta]\sigma}}^\nu=0$. Even more, taking into acount another reduction path inside the RC class, evaluation of MAG field equations coming from  \eqref{3-25} on the condition $Q_{\nu\lambda\rho}=0$ produces a different result
\begin{eqnarray}
\tau_{\mu\nu}=0
\,\, ,\,\, \label{3-46}
\end{eqnarray}
\begin{eqnarray}
D_\nu {F_\alpha}^{\beta\sigma\nu}+\big(-{\delta^\sigma}_\mu{T^\lambda}_{\lambda\nu}+\frac{1}{2}\,{T^\sigma}_{\mu\nu}\big){F_\alpha}^{\beta\mu\nu}=0
 \,\, ,\,\, \label{3-47}
\end{eqnarray}
whose  torsionless limit leads to a connection field equation in the form $\nabla_\nu {F_{\alpha}}^{\beta\sigma\nu}=0$.

Another posible reduction path in the RC class consist in to incorporate the  non-metricity through the action
\begin{eqnarray}
S''=S^{(n)}+k^{2n-N}\, \langle a^{\alpha\mu\nu}Q_{\alpha\mu\nu}\,\rangle
\,\, ,\,\, \label{3_48}
\end{eqnarray}
where $a^{\lambda\alpha\beta}=a^{\lambda\beta\alpha}$. The field equations are   
\begin{eqnarray}
\tau_{\mu\nu}-D_\alpha {{E_{\mu}}^\alpha}\,_{\nu}+{T^\alpha}_{\alpha\sigma} {{E_{\mu}}^\sigma}\,_{\nu}=0
\,\, ,\,\, \label{3-49}
\end{eqnarray}
\begin{eqnarray}
E_{[\mu\sigma\nu]}=0
 \,\, ,\,\, \label{3-50}
\end{eqnarray}
\begin{eqnarray}
a^{\nu\lambda\sigma}=-E^{\{\lambda\nu\sigma\}}
 \,\, ,\,\, \label{3-51}
\end{eqnarray}
coinciding con \eqref{3-43} and \eqref{3-44} in addition to obtaining non-trivial multipliers. 
Thus, we see that the procedure with the Lagrange multipliers, either in the EW or RC scenario, reproduces what we could call the expected field equations (but not those of a pseudo-Riemannian manifold), in contrast to what happens when the MAG expressions are specialized for each space-time. This will be evidenced once again in the next section.

\subsubsection{Riemann-Christoffel class (RCh)}

First we will review Hilbert's variational principle for \eqref{3-20} and for this we consider the functional variation of the connection in a pseudo-Riemannian space, that is
\begin{eqnarray}
\delta{A^\rho}_{\alpha\beta}\mid_{T=Q=0}\,\,=\delta{\Gamma^\rho}_{\alpha\beta}=\frac{g^{\rho\sigma}}2\,(\nabla_     
\alpha \delta g_{\sigma\beta}+\nabla_\beta \delta g_{\sigma\alpha}-\nabla_\sigma \delta g_{\alpha\beta})
\,\, ,\,\, \label{3-52}
\end{eqnarray}
and from \eqref{3-25} we write
\begin{eqnarray}
\delta \big( S^{(n)}\mid_{Q=T=0}\big) = k^{2n-N} \, \langle \big(\tau_{\mu\nu}
-\nabla_\sigma {{E_{\{\mu}}^\sigma}\,_{\nu\}}-\nabla_\sigma {E_{\{\mu\nu\}}}^\sigma +\nabla_\sigma {E^\sigma}_{\{\mu\nu\}}\big)\delta g^{\mu\nu}\rangle 
\,\, ,\,\, \nonumber \\
\label{3-53}
\end{eqnarray}
from where the following field equation emerges
\begin{eqnarray}
\tau_{\mu\nu}+\nabla_\sigma \big({E^\sigma}_{\{\mu\nu\}} -{{E_{\{\mu}}^\sigma}\,_{\nu\}}-{E_{\{\mu\nu\}}}^\sigma \big)=0
\,\, ,\,\, \label{3-54}
\end{eqnarray}
where, obviously all objects are evaluated on null torsion and non-metricity. This field equation will be our reference to compare with the reduction paths from the MAG. We have seen that none of the reduction classes (EW or RC) reproduces \eqref{3-54}, only in the case of the curvature function of order $n=1$ this is guaranteed. Recall that in the case in which we evaluate the MAG field equations (obtained from \eqref{3-25}) on the null torsion and non-metricity condition, that is $ \tau_{\mu \nu} = 0 $ and $ \nabla_\nu {F_\alpha}^{\beta\sigma\nu} = 0 $, they do not reproduce \eqref{3-54}. However, by introducing the torsion and non-metricity constraints, we write
\begin{eqnarray}
S'''=S^{(n)}+k^{2n-N}\, \langle a^{\alpha\mu\nu}Q_{\alpha\mu\nu}+{b_\alpha}^{\mu\nu}{T^\alpha}_{\mu\nu}\,\rangle
\,\, ,\,\, \label{3-55}
\end{eqnarray}
from which the following equations of motion are obtained
\begin{eqnarray}
\tau_{\mu\nu}+\nabla_\sigma \big({E^\sigma}_{\{\mu\nu\}} -{{E_{\{\mu}}^\sigma}\,_{\nu\}}-{E_{\{\mu\nu\}}}^\sigma \big)=0
\,\, ,\,\, \label{3-60}
\end{eqnarray}
\begin{eqnarray}
a_{\beta\mu\alpha}=-E_{\{\alpha\mu\}\beta} -E_{\{\alpha\beta\mu\}}+E_{\beta\{\mu\alpha\}}
\,\, ,\,\, \label{3-61}
\end{eqnarray}
\begin{eqnarray}
b_{\alpha\beta\mu}=E_{\alpha[\beta\mu]}+E_{[\beta\mu]\alpha}+E_{[\beta\alpha\mu]} 
\,\, ,\,\, \label{3-62}
\end{eqnarray}
\begin{eqnarray}
{T^\sigma}_{\mu\nu} =0
\,\, ,\,\, \label{3-58}
\end{eqnarray}
\begin{eqnarray}
Q_{\alpha\mu\nu}=0
\,\, ,\,\, \label{3-59}
\end{eqnarray}
clearly reproducing the field equation in a pseudo-Riemannian manifold.

Note that for the purposes of the presented discussion we do not require the introduction of external sources. However, it could be considered a term with external fields represented by some ''$\psi$'' in the style of the  reference\cite{Hehl2} and whose action could also depends on the metric and the affine connection (without considering the possible breaking of the projective symmetry), this means
\begin{eqnarray}
S^{(n)}=k^{2n-N}\, \langle F^{(n)}(R) +\frac{1}{\kappa_0}L(g,A,\psi)\,\rangle \,\, .\,\, \label{3-63}
\end{eqnarray}
and its variation is
\begin{eqnarray}
\delta S^{(n)} =k^{2n-N}\,\langle {\tau}'_{\mu\nu}\delta g ^{\mu\nu}
-2{E'_\alpha}^{\mu\beta}\delta{A^\alpha}_{\mu\beta}\rangle
\,\, ,\,\, \label{3-64}
\end{eqnarray}
with
\begin{eqnarray}
{\tau}'_{\mu\nu}=\tau_{\mu\nu}
-\frac{1}{2\kappa_0}\,T_{\mu \nu} \,\, ,\,\, \label{eq5}
\end{eqnarray}
\begin{eqnarray}
{E'_\alpha}^{\mu\beta}={E_\alpha}^{\mu\beta}-\frac{1}{2\kappa_0}\,{J_\alpha}^{\mu\beta}
\,\, ,\,\, \label{eq6}
\end{eqnarray}
where $T_{\mu \nu} \equiv \frac{-2\kappa_0
k}{\sqrt{-g}}\,\frac{\delta L}{\delta
g^{\mu\nu}}$ and ${J_\alpha}^{\mu\beta} \equiv
\frac{\kappa_0 k}{\sqrt{-g}}\,\frac{\delta L}{\delta
{A^\alpha}_{\mu\beta}}$ are the energy-momentum tensor of material fields and the hypermomentum current\cite{Hehl}, respectively. Particularly in the case $ n = 1 $ the Lagrange multipliers are no longer trivially zero and would depend exclusively on the hypermomentum current.

\section{Metric-affine $F^{(2)}(R)$ model}

The main purpose here is to explore the existence of vacuum solutions in the context of the study of reduction from a MAG manifold and how this affects the appropriate shape of the Lagrangian density. At the pseudo-Riemannian limit, where torsion and non-metricity are null, we will consider the Schouten-Weyl theorem as a key element to identify possible conformally flat solutions. From this, our interest will be focused on looking at models of the class $ 
n=2$. In a non-pseudo-Riemannian $N$-dimensional manifold, the most general possible model $ F^{(2)}(R)$ without sources is
\begin{eqnarray}
F^{(2)}(R)=a_1\,R^{\mu\nu\alpha\beta}R_{\mu\nu\alpha\beta}
+a_2\,R^{\mu\nu\alpha\beta}R_{\mu\alpha\nu\beta}+a_3\,R^{\mu\nu\alpha\beta}R_{\alpha\nu\mu\beta}
+a_4\,R^{\mu\nu\alpha\beta}R_{\alpha\mu\nu\beta}+\nonumber \\
+\,a_5\,R^{\mu\nu\alpha\beta}R_{\nu\mu\alpha\beta}
+a_6\,R^{\mu\nu\alpha\beta}R_{\alpha\beta\mu\nu}+a_7\,R^{\mu\nu}R_{\mu\nu}
+a_8\,R^{\mu\nu}R_{\nu\mu} +a_9\,R^2
\,\, ,\,\, \nonumber \\
\label{4-1}
\end{eqnarray}
and its explicit derivative with respect to the Riemann tensor is
\begin{eqnarray}
{F_\alpha}^{\beta\mu\nu}\equiv \frac{\partial F^{(2)}(R)}{\partial
{R^\alpha}_{\beta\mu\nu}}=2a_1\,{R_{\alpha}}^{\beta\mu\nu}
+2a_2\,{R_{\alpha}}^{[\mu\beta\nu]}+2a_3\,{{R^{[\mu\beta}}_\alpha}\,^{\nu]}
+a_4\,({{R^{\beta[\mu}}_\alpha}\,^{\nu]}+\nonumber \\+{{R^{[\mu}}_{\alpha}}\,^{\beta\nu]})
+2a_5\,{{R^{\beta}}_{\alpha}}^{\mu\nu}
+2a_6\,{{R^{[\mu\nu]}}_{\alpha}}\,^{\beta}+2a_7\,R^{\beta[\nu}{\delta^{\mu]}}_{\alpha}
+2a_8\,R^{[\nu\beta}{\delta^{\mu]}}_{\alpha}+\nonumber \\
+\,2a_9\,R\,g^{\beta[\nu}{\delta^{\mu]}}_{\alpha} \,\, ,\,\,\nonumber \\
\label{4-2}
\end{eqnarray}
its evaluation in a pseudo-Riemannian space is
\begin{eqnarray}
{F_\alpha}^{\beta\mu\nu}\big|_{T=Q=0}=a\,({{R^{\nu\beta}}_\alpha}\,^{\mu}-
{{R^{\nu}}_{\alpha}}\,^{\beta\mu})+b\,(R^{\nu\beta}{\delta^{\mu}}_{\alpha}-R^{\mu\beta}{\delta^{\nu}}_{\alpha})
+ \nonumber \\+c\,R\,(g^{\beta\nu}{\delta^{\mu}}_{\alpha}-g^{\beta\mu}{\delta^{\nu}}_{\alpha})
\,\, ,\,\,
\label{4-3}
\end{eqnarray}
with parameters
\begin{eqnarray}
a\equiv -2a_1-a_2-a_3 +a_4 +2a_5-2a_6 \,\, ,\,\, \label{4-4}
\end{eqnarray}
\begin{eqnarray}
b\equiv a_7+a_8 \,\, ,\,\, \label{4-5}
\end{eqnarray}
\begin{eqnarray}
c\equiv a_9 \,\,.\,\, \label{4-6}
\end{eqnarray}
Let us also remember that with null torsion and non-metricity conditions, the Bianchi identities are of the form
\begin{eqnarray}
\nabla_{\alpha}R_{\mu\nu\beta\lambda}+\nabla_{\lambda}R_{\mu\nu\alpha\beta}+\nabla_{\beta}R_{\mu\nu\lambda\alpha}=0
\,\, ,\,\, \label{4-7}
\end{eqnarray}
and all the well-known symmetry properties of the Riemann-Christoffel tensor, which will be useful when we examine the various reduction classes MAG$\rightarrow$RCh of $F^{(2)}(R)$ theories. With the help of the Bianchi identities and their contractions, we can write an identity in a pseudo-Riemannian space that we will eventually use
\begin{eqnarray}
\nabla_\nu {F_\alpha}^{\beta\mu\nu}\equiv (a-b)\nabla_\alpha R^{\beta\mu}-a \nabla^\beta{R_\alpha}^\mu+ (c+\frac{b}2){\delta^\mu}_\alpha \nabla^\beta R-cg^{\beta\mu}\nabla_\alpha R\,\,.\,\, \label{4-8}
\end{eqnarray}

\subsubsection{$F^{(2)}(R)$ and the Einstein-Weyl Class.}

Here we attempt the reduction to a pseudo-Riemannian space or RCh via Einstein-Weyl (i.e., ${T^\alpha}_{\mu\nu}=0 $, see fig.1) using the system \eqref{3-30} and \eqref{3-31} for the particular case $ n=2 $. From \eqref{4-3} we can write the pseudo-Riemannian limit of the connection equation as ${{E^{(T=0)}}_\alpha}^{\{\mu\beta\}}\big|_{Q=0}=\nabla_\nu{F_\alpha}^{\{\beta\mu\}\nu}=0 $ and with the help of the identity \eqref{4-8}, we can establish it in terms of first derivatives of the Ricci tensor as follows
\begin{eqnarray}
0={{E^{(T=0)}}_\alpha}^{\{\mu\beta\}}\big|_{Q=0}\equiv &-\frac{a}2 \nabla^\beta {R_\alpha}^\mu -\frac{a}2 \nabla^\mu {R_\alpha}^\beta +(a-b)\nabla_\alpha R^{\beta\mu}+\nonumber \\ 
&\frac{1}2 (c+\frac{b}2)({\delta^\mu}_\alpha \nabla^\beta R +{\delta^\beta}_\alpha \nabla^\mu R) -cg^{\beta\mu}\nabla_\alpha R \,\,.\,\, \label{4-9}
\end{eqnarray}

With the objective to explore existence of conformally flat solutions coming from ${{E^{(T=0)}}_\alpha}^{\{\mu\beta\}}\big|_{Q=0}=0$, we use the next algebraic combination
\begin{eqnarray}
0=\dfrac{1}{N-2}\,{E^{(T=0)}}_{\alpha\{\mu\beta\}}\big|_{Q=0} -\dfrac{1}{(N-2)(N-1)}\,\Big(g_{\mu\alpha}\,{{E^{(T=0)}}_{\lambda\{\beta}}^{\lambda\}}\big|_{Q=0} \nonumber \\
+\,g_{\alpha\beta}\,{{E^{(T=0)}}_{\lambda\{\mu}}^{\lambda\}}\big|_{Q=0}-2 g_{\mu\beta}\,{{E^{(T=0)}}_{\lambda\{\alpha}}^{\lambda\}}\big|_{Q=0}\Big)=\nonumber \\
= a \big(\nabla_{[\alpha} \mathcal{R}_{\beta]\mu} +\nabla_{[\alpha} \mathcal{R}_{\mu]\beta} \big)-\dfrac{b}{N-2}\,\nabla_\alpha G_{\mu\beta}\,\,,\,\, \label{4-12}
\end{eqnarray}
 where $\mathcal{R}_{\alpha\beta}$ and $G_{\alpha\beta}$ are the Schouten's and Einstein's tensors defined by
\begin{eqnarray}
\mathcal{R}_{\alpha\beta}\equiv\frac{1}{N-2}\big(R_{\alpha\beta}-\frac{g_{\alpha\beta}}{2(N-1)}\,R\big)
\,\, ,\,\, \label{4-13}
\end{eqnarray}
and
\begin{eqnarray}
G_{\alpha\beta}\equiv R_{\alpha\beta}-\frac{g_{\alpha\beta}}2\,R
\,\, ,\,\, \label{4-14}
\end{eqnarray}
respectively.
Equation \eqref{4-12} reveals a possible deviation from Shouten-Weyl's condition which stablishes conformally flatness because the Schouten tensor is not guaranteed to be of the Codazzi type, in other words, in general $\nabla_{[\alpha} \mathcal{R}_{\beta]\mu}\neq 0$. In order to look at this more precisely, we seek to remove the explicit presence of the Einstein's tensor in \eqref{4-12}, assuming $a\neq 0$ and $b\neq 0$. Working with symmetric-antisymmetric parts of \eqref{4-12} and with the help of definitions for Schouten and Einstein tensors, we get
\begin{eqnarray}
(3a-2b) \nabla_{[\alpha} \mathcal{R}_{\beta]\mu}= \dfrac{b}{N-1}\,g_{\mu[\alpha}\nabla_{\beta]} R \,\,,\,\, \label{4-17}
\end{eqnarray}
and
\begin{eqnarray}
0= b( \nabla_{\alpha} G_{\beta\mu}+\nabla_{\beta} G_{\mu\alpha}+\nabla_{\mu} G_{\alpha\beta}) \,\,.\,\, \label{4-18}
\end{eqnarray}
Furthermore, the trace of \eqref{4-18} is $0=-\dfrac{b(N-2)}2 \,\nabla_\mu R $, and with this the system of equations is
\begin{eqnarray}
(3a-2b) \nabla_{[\alpha} R_{\beta]\mu}= 0 \,\,,\,\, \label{4-19}
\end{eqnarray}
\begin{eqnarray}
b(\nabla_{\alpha} R_{\beta\mu}+\nabla_{\beta} R_{\mu\alpha}+\nabla_{\mu} R_{\alpha\beta})=0 \,\,,\,\, \label{4-20}
\end{eqnarray}
\begin{eqnarray}
b\nabla_\mu R= 0 \,\,,\,\, \label{4-21}
\end{eqnarray}
which admits  one solution of a cosmological type at least. However, in the particular case where the parameters satisfy
\begin{eqnarray}
3a-2b= 0 \,\,,\,\, \label{4-22}
\end{eqnarray}
the conformal flatness condition from \eqref{4-19} is no longer guaranteed and the constraints on the Ricci tensor are reduced to \eqref{4-20} and \eqref{4-21}. If one wishes to contrast \eqref{4-20} with the Schouten-Weyl condition for the case \eqref{4-21}, that is $\nabla_{[\alpha} R_{\beta]\mu} = 0$ it can be start by looking at the left member of \eqref{4-20} as a totally symmetric rank $3$ object and, considering that the indices take $N$ possible values, the number of independent components corresponds to that of combinations with repetition of $N$ elements in groups of $3$, it means
\begin{eqnarray}
{N+p-1 \choose p}\Big|_{p=3}= \frac{(N+2)(N+1)N}6 \,\,,\,\, \label{4-23}
\end{eqnarray}
which is the number of the mentioned combinations.
This indicates that the difference between the number of Schouten-Weyl conditions and \eqref{4-23} is
\begin{eqnarray}
\mathcal{D}(N)\equiv \frac{N^2(N-1)}2 - \frac{(N+2)(N+1)N}6\equiv \dfrac{N}3\,(N^2-3N-1) \,\,.\,\, \label{4-24}
\end{eqnarray}
In  $N=3$ dimension, $\mathcal{D}(3)=-1$, whereby there would be given redundant information, in addition to that the Weyl conformal tensor is identically null. However, in higher dimensions $N>3$ we have $ \mathcal{D}(N)>0$ in a monotonic way, which does not guarantee only conformally planar solutions under the condition \eqref{4-22}. Thus, we see that the existence of physically acceptable solutions is sensitive to the nullity or not of the factor $3a-2b$.

We have examined the connection equation and saw that there are configurations of the parameters that do not guarantee the existence of unique conformally flat solutions for the vacuum. This class of non-Einsteinian solutions can also be characterized through the equation of the metric, \eqref{3-30} for $n=2$, that is
\begin{eqnarray}
\tau_{\mu\nu}\equiv \dfrac{\partial F^{(2)}}{\partial g^{\mu\nu}}-\frac{F^{(2)}}{2}\,g_{\mu\nu}=0 \,\,,\,\, \label{4-25}
\end{eqnarray}
which can be  particularized for the case of the pseudo-Riemannian limit. With this, from \eqref{4-1} we write
\begin{eqnarray}
\tau_{\mu\nu}\Big|_{Q=T=0}=2(a_1-a_5+a_6)\,R_{\mu\alpha\beta\lambda}{R_\nu}^{\alpha\beta\lambda}
+2(a_2+a_3-a_4)\,R_{\mu\alpha\beta\lambda}{R_\nu}^{\beta\alpha\lambda}+\nonumber \\
+2(a_7+a_8)\,R_{\mu\lambda}{R_\nu}^\lambda +2a_9\,R R_{\nu\mu} -\dfrac{g_{\mu\nu}}2\,\Big[(a_1-a_5+a_6)\,R^{\alpha\beta\lambda\rho}R_{\alpha\beta\lambda\rho}+\nonumber \\ 
+(a_2-a_3-a_4)\,R^{\alpha\beta\lambda\rho}R_{\alpha\lambda\beta\rho}
+(a_7+a_8)\,R^{\alpha\beta}R_{\alpha\beta}+a_9\,R^2\Big]=0 \,\, .\,\,\nonumber \\ \label{4-26}
\end{eqnarray}

As an illustration we can try a very simple non-Einsteinian kind of solution thinking of a deviation with respect to the cosmological solution, i.e.
\begin{eqnarray}
\tilde{R}_{\mu\nu\lambda\rho} = \phi (x)\, (g_{\mu\lambda}g_{\nu\rho}-g_{\mu\rho}g_{\nu\lambda}) \,\, ,\,\, \label{4-27}
\end{eqnarray}
where $\phi (x)$ is an arbitray scalar field. By replacing \eqref{4-27} in \eqref{4-26} we get
\begin{eqnarray}
\tau_{\mu\nu}\Big|_{Q=T=0, \phi}=\dfrac{(N-1)}2\,q(a,b,c)\,\phi^2 g_{\mu\nu} =0\,\, ,\,\, \label{4-28}
\end{eqnarray}
where we have defined
\begin{eqnarray}
q(a,b,c)\equiv (N-4)[a-(N-1)b-N(N-1)c]\,\, ,\,\, \label{4-29}
\end{eqnarray}
and we see that \eqref{4-28} at $N=4$ is identically satisfied, leaving undetermined parameter $c$ and scalar field, $\phi$. However, in $N\neq 4$ dimensions, equation \eqref{4-28} tells us that under the condition $ q(a,b,c)=0$ it is possible to find this class of non-Einsteinian solutions (this last condition on the parameters together \eqref{4-22} allows to get an equivalent family of theories characterized only by the parameter $c$).

From the point of view of the connection equation, eq. \eqref{4-9}, when evaluating it on the solution \eqref{4-27} in a pseudo-Riemannian space, we write
\begin{eqnarray}
\big({{E^{(T=0)}}_\alpha}^{\{\mu\beta\}}\big|_{Q=0}\big)\big|_\phi=\dfrac{q(a,b,c)}{N-4}\,(g^{\{\beta \nu}{\delta^{\mu\}}}_\alpha -g^{\beta \mu}{\delta^\nu}_\alpha)\,\partial_\nu \phi =0\,\,,\,\, \label{4-30}
\end{eqnarray}
confirming that for the critical value $\dfrac{q(a,b,c)}{N-4}=0$ the scalar field remains undetermined, while for $\dfrac{q(a,b,c)}{N-4}\neq 0 $, the trace of \eqref {4-30} is $(1-N)\partial_\alpha \phi = 0$ , in a consistent way with the aforementioned observation about the metric equation \eqref{4-28}.

\subsubsection{$F^{(2)}(R)$ and the Riemann-Cartan Class.}

Starting with the equation of the connection  \eqref{3-44}, that is to say, $E_{[\alpha\sigma\beta]}\equiv D_\nu {F_{[\alpha\beta]\sigma}}^\nu-\big(g_{\sigma\mu}\,{T^\lambda}_{\lambda\nu}-\frac{1}{2}\,T_{\sigma\mu\nu}\big){F_{[\alpha\beta]}}^{\mu\nu}=0$, it can be especialized for $n=2$, then using identity \eqref{4-8}, at the torsionless limit an equivalent combination in terms of Schouten's tensor is written as follows
\begin{eqnarray}
{E^{(Q=0)}}_{[\alpha\mu\beta]}\big|_{T=0} -\dfrac{2}{N-1}\, g_{\mu\big[\alpha} {{{{E^{(Q=0)}}_{[\lambda}}^\lambda}\,_{\beta]}}_{\big]} \big|_{T=0} = \nonumber \\
=(2a-b)(N-2)\,\nabla_{[\alpha}  \mathcal{R}_{\beta]\mu}=0  \,\, ,\,\, \label{4-33}
\end{eqnarray}
being able to observe that  the condition for conformal flatness is not well defined in the critical value 
\begin{eqnarray}
2a-b=0 \,\, ,\,\, \label{4-33a}
\end{eqnarray}

Finally we look at some common quadratic theories
\begin{eqnarray}
L_{Yang-Mills}= -\dfrac{1}{4}\,R^{\mu\nu\alpha\beta}R_{\nu\mu\alpha\beta} \,\, ,\,\, \label{4-34a}
\end{eqnarray}
\begin{eqnarray}
L_{Gauss-Bonnet}= R^{\mu\nu\alpha\beta}R_{\mu\nu\alpha\beta}-4R^{\mu\nu}R_{\mu\nu}+ R^2\,\, ,\,\, \label{4-34b}
\end{eqnarray}
\begin{eqnarray}
L_{Weyl}= R^{\mu\nu\alpha\beta}R_{\mu\nu\alpha\beta}-2R^{\mu\nu}R_{\mu\nu}+ \dfrac{1}3\,R^2\,\, ,\,\, \label{4-34c}
\end{eqnarray}
and we order the parameters as follows
\begin{table}[htb]
\begin{center}
\begin{tabular}{| c | c | c | c | c |} 
\hline
 Theory  & parameters & $3a-2b$ & $2a-b$ & No-Einstein $-q(a,b,c)$\\ 
\hline
Y-M & $a=-1\diagup2,\,b=c=0$& $-3\diagup2$ & $-1$ & $(N-4)/2$ \\
\hline
G-B &$a=-2,\,b=-4,\,c=1$  & $2$ & $0$& $(N-2)(N-3)(N-4)$ \\
\hline
W &$a=-2,\,b=-2,\,c=1\diagup3$  & $-2$ & $-2$ &$(N-3)(N-4)^2/3$ \\
\hline
\end{tabular}
\end{center}
\end{table}

It can be observed that the Yang-Mills gravity type formulation does not present any problem (in any dimension) regarding the existence of conformally flat solutions, regardless of the path or  reduction class that is chosen from the MAG. While in the reduction of the Gauss-Bonnet action via the Riemann-Cartan class, the Schouten-Weyl conformal flatness condition is not explicitly guaranteed, which leaves open the possibility of non-conformally flat solutions at least in $ N = 3, \, 4 $ dimensions. Something similar occurs for the Weyl action as in the case of Yang-Mills regarding the existence of the expected vacuum solutions.

\section{The topologically massive gravity.}

The interest about the study of TMG has had several revivals both in the context of quantum gravity \cite{Witten} \cite{RovelliSmolin}, as well as from the point of view of the extension of the geometry in which this theory has been conceived. The latter occurs, for example, in the cases of the Mielke-Baekler formulation\cite{Mielke} for a Riemann-Cartan geometry and as in the case of the extension to spaces with non-zero torsion and non-metricity\cite{Tresguerres}. This spirit will be that of our fundamental interest. To this end, we will begin with a brief review on the physical content of the original TMG model\cite{DJT} and then explore its direct extension without major additives to the set of objects already established towards Riemann-Cartan and Einstein-Weyl spaces.

In the case of $2+1$ dimensional gravity, there are two remarkable non-perturbative formulations from which it is possible to obtain physical degree of
freedom (unitary massive gravitons). On the one hand, it is known the BHT formulation\cite{BHT}, which is
a modified Hilbert-Einstein theory plus particular quadratic terms in curvature, leading to
 a  fourth-order derivative metric action. This theory at its perturbative limit
reveals a physical content analogous to the Fierz-Pauli model. On the other hand, there is the
formulation of the TMG which in its original formulation consists in a
Hilbert-Einstein plus another of the Chern-Simons type based on the Lorentz connection and parity variant. Here we will focus
our attention in the TMG but from the point of view of general relativistic covariance, for which we consider
the Hilbert-Einstein terms and the topological Chern-Simons term as an explicit function of the
Levi-Civita connection, this means
\begin{equation}
S_{TMG} = -\sigma \langle R(\Gamma) \rangle -\frac{\sigma}{2m} \langle \varepsilon^{\mu\nu\lambda}( \Gamma^{\alpha}{_{\mu\beta}} \partial_\nu \Gamma^{\beta}{_{\lambda\alpha}} + \dfrac{2}{3}\Gamma^{\alpha}{_{\mu\beta}} \Gamma^{\beta}{_{\nu\gamma}}\Gamma^{\gamma}{_{\lambda\alpha}} )    \rangle \,\,\,,\label{5-1}
\end{equation}
where $ \varepsilon^{\mu\nu\alpha} = \frac{\epsilon^{\mu\nu\alpha}}{\sqrt{-g}} $ is the Levi-Civita tensor, with $ \epsilon^{012} = +1 $ and $\sigma$ and $m$ in $(lenght)^{-1}$ units.

Next we focus on the analysis of a cubic order system of constraints with a pedagogical purpose. The perturbation around a Minkowski space-time considers the metric as
\begin{equation}
g_{\mu\nu} = \eta_{\mu\nu} + h_{\mu\nu} \,\,\,, \,\,\, \,\,\, \,\,\,\mid h_{\mu\nu}\mid <<1 \,\,\,,\label{5-2}
\end{equation}
and $g^{\mu\nu} = \eta^{\mu\nu} - h^{\mu\nu}$, until first order. The linearized TMG action is of the form $S^{L}{_{GTM}}=S^{L}{_{HE}}+S^{L}{_{CS}}$ and give rise the following second order equation
\begin{equation}
\square h^{\alpha\beta} - \partial^{\alpha}\partial_\lambda h^{\lambda\beta} - \partial^{\beta}\partial_\lambda h^{\lambda\alpha} + \partial^{\alpha}\partial^{\beta} h + \frac{1}{m}\,\epsilon^{\{\beta\mu\nu}\partial_\mu (\square {h^{\alpha\}}}_\nu - \partial^{\alpha\}}\partial_\lambda {h^\lambda}_\mu) = 0 \,\,\,,\label{5-6} 
\end{equation}
which is invariant under the transformation 
\begin{equation}
\delta h_{\mu\nu} = \partial_\mu \xi_\nu + \partial_\nu \xi_\mu \,\,\,, \label{5-5}
\end{equation}
with an arbitrary vector $\xi_\nu$.  \eqref{5-5} can be obtained through a diffeomorphism transformation of a rank $2$ tensor or, equivalently performing the gauge symmetry study on the constrained Hamiltonian\cite{Cast}.

Lets start with the Lagrangian constraints analysis considering the Coulomb gauge fixation and its preservation 
\begin{equation}
\partial_i h_{i\mu} = 0 \,\,\,,\label{5-7} 
\end{equation}
\begin{equation}
\partial_i \dot{h}_{i\mu} = 0 \,\,\,.\label{5-7b} 
\end{equation}
With this in mind, we perform a $2+1$ decomposition in \eqref{5-6} in the way
\begin{equation}
\ddot{h}_{ii} + \Delta h_{00}   - \dfrac{1}{m}\epsilon_{ij}\Delta \partial_i h_{0j}=0\,\,, \label{5-8}
\end{equation}
\begin{align}
- \dfrac{1}{2m}\epsilon_{nm} \partial_n \ddot{h}_{jm}+ \dfrac{1}{2m}\epsilon_{nm}\partial_n \partial_j \dot{h}_{0m}  - \partial_j \dot{h}_{ii}- \dfrac{1}{2m}\epsilon_{nj} \Delta \dot{h}_{0n} +\nonumber \\
+ \dfrac{1}{2m} \epsilon_{nm} \partial_n \Delta h_{jm} -\Delta h_{0j}  
-\dfrac{1}{2m}\epsilon_{jn}\partial_n \Delta h_{00}  = 0 \,\,,\label{5-9}
\end{align}
\begin{align}
-\ddot{h}_{ij} + \Delta h_{ij} + \partial_i \dot{h}_{0j} + \partial_j \dot{h}_{0i} - \partial_i \partial_j h_{00} &+ \partial_i  \partial_j h_{nn} +\dfrac{1}{m}\epsilon_{n\{i}(-\dddot{h}_{j\}n}+\nonumber\\ + \Delta \dot{h}_{j\}n} + \partial_{j\}} \ddot{h}_{0n})+ 
&+\dfrac{1}{m} \epsilon_{\{in}\partial_n(-\ddot{h}_{j\}0} + \Delta h_{j\}0} + \partial_{j\}} \dot{h}_{00}) = 0 \,\,.\label{5-10}
\end{align}
Equations \eqref{5-8}, \eqref{5-9} and \eqref{5-10} constitute a cubic system in time derivatives and we are looking for constraints. It can be noted the longitudinal part of \eqref{5-9} is a constraint
\begin{equation}
\dot{h}_{nn} - \dfrac{1}{m} \epsilon_{nm}\partial_n \dot{h}_{0m} = 0 \,\,, \label{5-11}
\end{equation}
while the transverse part gives
\begin{equation}
-\ddot{h}_{nn} + \Delta h_{nn} -2m\epsilon_{nm} \partial_n h_{0m} + \Delta h_{00} = 0\,\,. \label{5-12}
\end{equation}
Even though this is not manifestly a Lagrangian constraint, when combining it with \eqref{5-8} we have the constraint
\begin{equation}
2\Delta h_{00} + \Delta h_{nn} - 2m(1+\dfrac{\Delta}{2m^2})(\epsilon_{nm}\partial_n h_{0m}) = 0 \,\,.\label{5-13}
\end{equation}
Until now we have seen that the components \eqref{5-8} and \eqref{5-9} mean two Lagrangian constraints (in other words, \eqref{5-11} and \eqref{5-13}) plus a relation for accelerations (i. e., either of the two \eqref{5-8} or \eqref{5-12}).  

Now let's look at \eqref{5-10}, which contains three components. On the one hand its trace is$
-\ddot{h}_{nn} + 2\Delta h_{nn} - \Delta h_{00} -\dfrac{1}{m} \epsilon_{nm} \Delta \partial_n h_{0m} = 0$ and combination
with \eqref{5-8} give rise to the constraint
\begin{equation}
h_{nn} - \dfrac{1}{m}\epsilon_{nm}\partial_n h_{0m} = 0 \,\,,\label{5-14}
\end{equation}
where we observe that by preserving \eqref{5-14} we obtain \eqref{5-11}. This means that the set of relations \eqref{5-11}, \eqref{5-13} and \eqref{5-14} are only two constraints, that is, \eqref{5-14} and \eqref{5-13}, where the latter can be rewritten as
\begin{equation}
\Delta h_{00} - m^{2}h_{nn} = 0 \,\,.\label{5-15}
\end{equation}

preservation of \eqref{5-14} and \eqref{5-15} give rise to another two constraints
\begin{equation}
\dot{h}_{nn} - \dfrac{1}{m}\epsilon_{nm}\partial_n \dot{h}_{0m} = 0 \,\,,\label{5-16}
\end{equation}
\begin{equation}
\Delta \dot{h}_{00} - m^{2}\dot{h}_{nn} = 0\,\,, \label{5-17}
\end{equation}
whose preservations represent relations for accelerations. With the help of \eqref{5-8} it can be written the following accelerations 
\begin{equation}
\ddot{h}_{0i} = m\epsilon_{ri} \partial_r h \,, \label{5-18}
\end{equation}
\begin{equation}
\ddot{h}_{00} = m^2 h \,,\label{5-19}
\end{equation}
and $ \ddot{h}_{ij} $ remains undetermined. 

Using \eqref{5-18}, relation \eqref{5-10} is rewritten 
\begin{eqnarray}
\dfrac{1}{m}\epsilon_{\{in} \dddot{h}_{nj\}} - \ddot{h}_{ij} - \dfrac{1}{m} \epsilon_{\{in}\Delta \dot{h}_{nj\}} + 2\partial_{\{i} \dot{h}_{0j\}} + \dfrac{1}{m} \epsilon_{\{in} \partial_n \partial_{j\}} \dot{h}_{00} \nonumber \\+ \Delta h_{ij} - \partial_i \partial_j h 
+ \eta_{ij}\Delta +h\dfrac{1}{m} \epsilon_{\{in} \Delta \partial_n h_{0j\}} = 0 \,,\label{5-20}
\end{eqnarray}
which allow us to solve $ \dddot{h}_{ij} $ in terms of $ \dot{h}_{\mu\nu} $, in other words
\begin{equation}
\dddot{h}_{ij} = \Delta\dot{h}_{ij} - 2m \epsilon_{n\{i} \partial_{n}\dot{h}_{0j\}} - (\partial_i \partial_j - \eta_{ij}\Delta)\dot{h}_{00} \,\,,\label{5-21}
\end{equation}
which we insert in \eqref{5-20} to obtain the remaining accelerations
\begin{equation}
\ddot{h}_{ij} = \Delta h_{ij} + (\eta_{ij} - \partial_i \partial_j)h + \dfrac{1}{m} \epsilon_{\{in}\Delta \partial_n h_{0j\}} \,,\label{5-22}
\end{equation}
thus culminating the Lagrangian analysis.

In summary we have the following system of ten constraints
\begin{equation}
\partial_i h_{i\mu} = 0\,,\label{5-23} 
\end{equation}
\begin{equation}
\partial_i \dot{h}_{i\mu} = 0\,,\label{5-24}
\end{equation}
\begin{equation}
h_{nn} - \dfrac{1}{m}\epsilon_{nm}\partial_n h_{0m} = 0\,,\label{5-27}
\end{equation}
\begin{equation}
\Delta h_{00} - m^{2}h_{nn} = 0\,,\label{5-28}
\end{equation}
\begin{equation}
\dot{h}_{nn} - \dfrac{1}{m}\epsilon_{nm}\partial_n \dot{h}_{0m} = 0\,,\label{5-29} 
\end{equation}
\begin{equation}
\Delta \dot{h}_{00} - m^{2}\dot{h}_{nn} = 0\,,\label{5-30}
\end{equation}
and accelerations 
\begin{equation}
\ddot{h}_{0i} = m\epsilon_{ni}\partial_n h\,,\label{5-31}
\end{equation}
\begin{equation}
\ddot{h}_{00} = m^{2}h\,,\label{5-32}
\end{equation}
\begin{equation}
\ddot{h}_{ij} = \Delta h_{ij} + (\eta_{ij}\Delta - \partial_i \partial_j)h + \dfrac{1}{m} \epsilon_{\{in}\Delta \partial_n h_{0j\}}\,,\label{5-33}
\end{equation}
representing a single degree of freedom, as it should be.

\subsection{The topologically massive metric-affine gravity: symmetries.} 

In this section we will contrast what we have approached as a gauge symmetry on the bundle $GL(3,R)$ given in \eqref{ecu:11}
and the difeomorphisms symmetry of space (Appendix A). We will focus our attention on the behaviour of topological term
essentially because the boundary contributions. Let us remember that the model in question is described by the action
\begin{equation}
S_{TMAG} = S_0+ S_{CS} \,\,, \label{6-1}
\end{equation}	
where $S_0$ is the Hilbert-Einstein-metric-affine action given at \eqref{ecu:26} and $S_{CS}$ is the Chern-Simons action in terms of the affine connection, this means
\begin{equation}
S_{CS} = - \dfrac{\sigma}{2m} \langle \varepsilon^{\mu\nu\lambda} tr(A_\mu\partial _\nu A_\lambda + \dfrac{2}{3} A_\mu A_\nu A_\lambda) \rangle \,\,, \label{6-2}
\end{equation}	
under general coordinates transformation \eqref{ecu:1}, the last one changes as
\begin{eqnarray}
S'_{CS} = - \dfrac{\sigma}{2m} \langle \varepsilon'^{\mu\nu\lambda}( A'^{\alpha}{_{\mu\beta}} \partial'_\nu A'^{\beta}{_{\lambda\alpha}} + \dfrac{2}{3}A'^{\alpha}{_{\mu\beta}} A'^{\beta}{_{\nu\gamma}}A'^{\gamma}{_{\lambda\alpha}} )\rangle \nonumber \\
= - \dfrac{\sigma}{2m} \langle \varepsilon^{\mu\nu\lambda}tr( A'_\mu \partial_\nu A'_\lambda + \dfrac{2}{3}A'_\mu A'_\nu A'_\lambda)\rangle\,, \label{6-3}
\end{eqnarray}
where the matrix representation is performed by $ (A'_{\mu})^{\alpha}{_{\beta}} \equiv U^{\nu}{_{\mu}} A'^{\alpha}{_{\nu\beta}} $ with $ U \in GL(3,R) $.
From \eqref{6-3} we can write the Chern-Simons term variation with respect to general transformations as
\begin{eqnarray}
\delta_U S_{CS}=
- \dfrac{\sigma}{2m}\langle \varepsilon^{\mu\nu\lambda} tr\partial_\nu(\partial_\mu U^{-1}U A_\lambda) \rangle + \nonumber \\ +\dfrac{\sigma}{6m}\langle \varepsilon^{\mu\nu\lambda} tr (U\partial_\mu U^{-1} U \partial_\nu U^{-1}U\partial_\lambda U^{-1}) \rangle \,\,. \label{6-4}
\end{eqnarray}
The expression \eqref{6-4}, equivalent to the already reported\cite {DJT}, reaffirms the variant character
of the Chern-Simons action. On the one hand, the first term of the right-hand side of \eqref{6-4} is a boundary term that it can be removed by imposing suitable asymptotic conditions on the elements of the group, for example
$ \lim \limits_{x\to \pm \infty}U=1$ and an bounded affine connection at the asymptotic limit. On the other hand the
second term of the right side of \eqref{6-4} is a bit more complicated due to the non-compact character
of $GL(3,R)$ and therefore we could only ensure its proportionality to a integer number (i.e., winding number)
restricting the general covariance to a compact subgroup like $SO(3)$ (locally isomorphic to $ SU(2)$) which is doubly connected so
the associated winding number is half that of $SU(2)$.

Now we look at the transformation under diffeomorphisms of $S_ {CS}$. A diffeomorphism is a smooth, continuous, differentiable and invertible deformation over space-time that induces the change on space-time coordinates given by  $ x'^{\mu} = x^{\mu} -\xi^{\mu}(x) $ with $\xi^{\mu}(x) $ continuous and differentiable. At the same time this induces a diffeomorphisms transformation on the fields as indicated in Appendix A. Then, at the first order in $ \delta \xi^\mu $ we have that the variation of $ S_{CS} $ is
\begin{equation}
\delta_\xi S_{CS}=- \dfrac{\sigma}{2m}\int d^{3}x \,\partial_\lambda\left( \epsilon^{\mu\nu\lambda}\partial_\mu {A^\alpha}_{\nu\beta}\partial_\alpha \xi^\beta +  \epsilon^{\mu\nu\sigma}l_{\mu\nu\sigma}\xi^{\lambda}\right)\,,
\label{6-5}
\end{equation}
where  $ l_{\mu\nu\lambda} \equiv tr(A_\mu \partial_\nu A_\lambda + \dfrac{2}{3}A_\mu A_\nu A_\lambda)$. It can be observed that the right-hand of  \eqref{6-5} is a boundary term no matter the non-compact character of $GL(3,R)$, in contrast with \eqref{6-4}. There is a non-local difference between gauge and diffeomorphisms transformations.

\section{The topologically massive gravity in a Riemann-Cartan space-time.}

In this section we want to review the physical content of the TMG in a Riemann-Cartan space which has already been
approached from the point of view of the coframe field and the connection 1-form  as dynamic objects\cite{Mielke}. As we have established since
at the beginning, our interest is to consider the decomposition of the affine connection in terms of torsion and non-metricity
according to the expression \eqref{ecu:15}, from which, in the Riemann-Cartan regime we will show that the torsion contains, among others, the
propagation of a spin $2$ graviton (torsionon) described by a modified selfdual theory that extends the one already known in 
reference\cite{Aragone}.

Then, in the context of a Riemann-Cartan space, we impose the null non-metricity condition and therefore we will have that the only
dynamical objects are the metric and the torsion (contortion) and from the expression \eqref{ecu:15} we get
\begin{equation}
A^{\lambda}{_{\mu\nu}} = \Gamma^{\lambda}{_{\mu\nu}}(g) + K^{\lambda}{_{\mu\nu}}(T) \,, \label{7-1}
\end{equation}
or in a matricial way, $ A_\mu = \Gamma_\mu + K_\mu $. Here we observe that obviously not all the components of $A_\mu$ are independent since the null non-metricity condition imposes restrictions on the symmetric part of the connection as indicated in \eqref{3-12}. The Riemann-Cartan tensor in its matrix form is
$\mathbf{R}_{\mu\nu} = \partial_\nu A_\mu - \partial_\mu A_\nu + \left[ A_\nu, A_\mu \right]$ or in terms of the Levi-Civita derivative it can be written as
\begin{equation}
\mathbf{R}_{\mu\nu} = \mathbf{R}_{\mu\nu}(\Gamma) + \left[K_\nu, \Gamma_\mu \right] +\nabla_\nu K_\mu-\nabla_\mu K_\nu \,, \label{7-2}
\end{equation}
where $ \mathbf{R}_{\mu\nu}(\Gamma) $ is the Riemann-Christofell curvature defined at \eqref{ecu:21}. From \eqref{7-2} is obtained the Ricci-Cartan tensor and scalar
\begin{equation}
{R}_{\beta\nu} \equiv {R^\alpha}_{\beta\alpha\nu} = {R}_{\beta\nu}(\Gamma) + \left[ K_\nu, K_\alpha \right]^{\alpha\beta}+ \nabla_\nu K^{\alpha}{_{\alpha\beta}}- \nabla_\alpha K^{\alpha}{_{\nu\beta}} \,, \label{7-3}
\end{equation}
\begin{equation}
R = R(\Gamma) + \left[ K_\nu,K_\alpha \right]^{\alpha\nu} + 2\nabla_\sigma (K^{\alpha}{_{\alpha}}{^{\sigma}} ) \,, \label{7-4}
\end{equation}
and the Hilbert-Einstein-Riemann-Cartan action is
\begin{equation}
S_{HE}[g,K] = -\sigma \langle R(\Gamma) + [K_\nu,K_\alpha]^{\alpha\nu} \rangle \,, \label{7-5}
\end{equation}
up to a boundary term.

Using \eqref{7-1} in Chern-Simons action, we write down
\begin{eqnarray}
S_{CS}[g,K] = S_{CS}(\Gamma) - \dfrac{\sigma}{2m} \langle \varepsilon^{\mu\nu\lambda} tr\,( K_\mu \nabla_\nu K_\lambda + \dfrac{2}{3}K_\mu K_\nu K_\lambda )  \rangle +\nonumber \\ - \dfrac{\sigma}{2m} \left\langle \varepsilon^{\mu\nu\lambda} tr K_\mu \mathbf{R}_{\lambda\nu}(\Gamma) \right\rangle \,\,,  \label{7-6}
\end{eqnarray}
where we see that the Chern-Simons-Riemann-Cartan action decomposes as a Chern-Simons-Riemann-Christoffel plus a tensor-Chern-Simons type term in the contortion and a third term that complements the coupling of contortion with the Riemann-Christoffel background.

To write the action of TMG in Riemann-Cartan (TMGRC) we will consider the fact that in $2+1$ dimensions the Riemann-Christoffel tensor can be written in exclusive terms of the Ricci tensor and scalar (i. e., $ R_{\lambda\mu\nu\sigma}(\Gamma) = g_{\lambda\nu}R_{\mu\sigma}(\Gamma) - g_{\lambda\sigma} R_{\mu\nu}(\Gamma) - g_{\mu\nu}R_{\lambda\sigma}(\Gamma) + g_{\mu\sigma}R_{\lambda\nu}(\Gamma) - 1/2\,R(\Gamma)(g_{\lambda\nu}g_{\mu\sigma} - g_{\lambda\sigma}g_{\mu\nu} )$) and also the contortion can be decomposed into symmetrical part (i. e., $h_{\sigma\rho}=h_{\rho\sigma}$) and antisymmetrical (i. e., $t^{\mu}$) as follows
\begin{equation}
K^{\mu}{_{\rho\nu}} \equiv \varepsilon^{\mu}{_{\nu}}{^{\sigma}}k_{\sigma\rho}= \varepsilon^{\mu}{_{\nu}}{^{\sigma}}h_{\sigma\rho} + g_{\rho\nu}t^{\mu} - \delta^{\mu}{_\rho} t_\nu \,, \label{7-7}
\end{equation}
where we have used
\begin{equation}
k_{\sigma\rho}= h_{\sigma\rho} + {\varepsilon_{\sigma\rho}}^\lambda  t_\lambda \,. \label{7-7a}
\end{equation}

Thus, considering \eqref{7-5} and \eqref{7-6} with the help of \eqref{7-7}, the TMGRC action is
\begin{eqnarray}
S_{TMGRC} = S_{TMG} [\Gamma] + \tilde{S}_{SD} [k]+S_{int} \,, \label{7-8}
\end{eqnarray}
where
\begin{equation}
S_{TMG}[\Gamma] = -\sigma \langle R(\Gamma) \rangle -\frac{\sigma}{2m} \langle \varepsilon^{\mu\nu\lambda}tr\,( \Gamma_\mu \partial_\nu \Gamma_\lambda + \dfrac{2}{3}\Gamma_\mu\Gamma_\nu\Gamma_\lambda)    \rangle \,, \label{7-9}
\end{equation}
is the TMG action in Riemann-Christoffel space, while
\begin{eqnarray}
\tilde{S}_{SD}[k] = \dfrac{\sigma}{m}\langle  -\varepsilon^{\mu\nu\lambda} k_{\sigma\mu} \nabla_\nu k^{\sigma}{_{\lambda}} - m (k_{\mu\nu }k^{\nu\mu} - k^{2}) + \nonumber \\- \dfrac{1}{3}k^{3} - \dfrac{2}{3}k_{\mu\nu} k^{\lambda\mu} k^{\nu}{_{\lambda}} + kk_{\mu\nu}k^{\nu\mu} \rangle \,, \label{7-10}
\end{eqnarray}
is proportional to a modified selfdual action with cubic self-interacion potencial. The last term of the right side of \eqref{7-8}, this means
\begin{eqnarray}
S_{int} = -\dfrac{2\sigma}{m}\langle h_{\mu\nu} G^{\mu\nu}(\Gamma) \rangle \,, \label{7-11} 
\end{eqnarray}
contributes to interaction ''graviton-torsionon'' where $G^{\mu\nu}(\Gamma)=R^{\mu\nu}(\Gamma)-\frac{g^{\mu\nu}}2\,R(\Gamma)$ is the Einstein's tensor.

While the action \eqref{7-9} represents a massive graviton according to the DJT model, our fundamental interest will be in to
explore the physical content of what we have called {\it modified selfdual} action, described in \eqref{7-10}. At the limit of
the weak torsion (contortion) or weak Weitzenb\"{o}ck regime, performing a perturbative analysis at second order in
fields and removing the self-interaction cubic potential, we would have a pure spin $2$ selfdual theory\cite{Aragone} \cite{Arias} that
propagates a single massive degree of freedom with spin $2$. Note that the sign of the kinematical Chern-Simons term is negative and does not
affect the sign of the Hamiltonian. In any case unitarity 
in the weak Weitzenb\"{o}ck regime depends on the sign of the Fierz-Pauli term, which is the correct one.

Thus, a first approach to the study of the effect of the cubic potential presented in \eqref{7-10}, we will consider a Minkowskian metric background in order to focus on the behavior of the contortion due to this kind of self-interaction. Let us then consider the action
\begin{eqnarray}
S\equiv \dfrac{m}{\sigma}\left. \tilde{S}_{SD}[k] \right|_{g_{\alpha\beta}=\eta_{\alpha\beta}}= \langle  -\epsilon^{\mu\nu\lambda} k_{\sigma\mu} \partial_\nu k^{\sigma}{_{\lambda}} - m (k_{\mu\nu }k^{\nu\mu} - k^{2}) + \nonumber \\- \dfrac{1}{3}k^{3} - \dfrac{2}{3}k_{\mu\nu} k^{\lambda\mu} k^{\nu}{_{\lambda}} + kk_{\mu\nu}k^{\nu\mu} \rangle \,, \label{7-11b}
\end{eqnarray}
and being its functional variation $\delta_k S= \langle -2 \Phi^{(1)\mu\nu} \delta k_{\nu\mu}\rangle$, the field equations constitute nine constraints
\begin{eqnarray}
0=\Phi^{(1)\mu\nu}\equiv \epsilon^{\mu\alpha\beta} \partial_\alpha {k^\nu}_\beta + m (k^{\mu\nu } - \eta^{\mu\nu }k) + {k^\mu}_\lambda k^{\lambda\nu} - kk^{\mu\nu} + \nonumber \\+\frac{\eta^{\mu\nu}}2\,(k^2 -k_{\alpha\beta}k^{\beta\alpha}) \,. \label{7-12}
\end{eqnarray}
One way to visualize the degrees of freedom of this theory will be done by looking at some algebraic properties of \eqref{7-12}. Since we are in $2+1$ dimensions, \eqref{7-12} is equivalent to
\begin{eqnarray}
\epsilon_{\mu\sigma\rho}\Phi^{(1)\mu\nu}\equiv \partial_\rho {k^\nu}_\sigma - \partial_\sigma {k^\nu}_\rho +\epsilon_{\mu\sigma\rho}[ m (k^{\mu\nu } - \eta^{\mu\nu }k) + {k^\mu}_\lambda k^{\lambda\nu} - kk^{\mu\nu} + \nonumber \\+\frac{\eta^{\mu\nu}}2\,(k^2 -k_{\alpha\beta}k^{\beta\alpha}) ] \,, \label{7-13}
\end{eqnarray}
and its trace provides the antisymmetric part of $\Phi^{(1)\mu\nu}$, in other words $\epsilon_{\sigma\mu\nu}\Phi^{(1)\mu\nu}\equiv \partial_\sigma k - \partial_\rho {k^\rho}_\sigma +\epsilon_{\sigma\mu\nu}[ m k^{\mu\nu }+ {k^\mu}_\lambda k^{\lambda\nu} - kk^{\mu\nu}]$ and using \eqref{7-7a} it can be written as
\begin{eqnarray}
\epsilon_{\sigma\mu\nu}\Phi^{(1)\mu\nu}\equiv \partial_\sigma k - \partial_\rho {k^\rho}_\sigma -2[m{\delta^\rho}_\sigma-{k^\rho}_\sigma]t_\rho \,. \label{7-14}
\end{eqnarray}

On the other hand, with the help of \eqref{7-13} and \eqref{7-14}, divergence  of \eqref{7-12} is
\begin{eqnarray}
\partial_\mu{\Phi^{(1)\mu}}_\nu \equiv -2m^2 t_\nu \,, \label{7-15}
\end{eqnarray}
saying that the antisymmetric part of $k_{\mu\nu}$ does not propagate degree of freedom (i. e., $k_{\mu\nu}=h_{\mu\nu}$) and considering \eqref{7-14}, we write the antisymmetric part of  \eqref{7-12} as follows
\begin{eqnarray}
\epsilon_{\sigma\mu\nu}\Phi^{(1)\mu\nu} \equiv \partial_\sigma h - \partial_\rho {h^\rho}_\sigma \,, \label{7-16}
\end{eqnarray}
stablishing that there is no spin $1$ propagation. Indeed using the TLt-decomposition of a symmetric rank $2$ tensor, this means $h_{\mu\nu}={h^{Tt}}_{\mu\nu} +\hat{\partial}_\mu {a^T}_\nu + \hat{\partial}_\nu {a^T}_\mu +\hat{\partial}_\mu\hat{\partial}_\nu \phi+\eta_{\mu\nu}\psi$ (Appendix B) in \eqref{7-16} it can be obtained ${a^T}_\mu=\psi=0$ and with this we see that the physical reduction on the original symmetric field arrives to
\begin{eqnarray}
h_{\mu\nu}(h^{Tt},\phi)={h^{Tt}}_{\mu\nu}  +\hat{\partial}_\mu\hat{\partial}_\nu \phi\,. \label{7-17}
\end{eqnarray}

Inserting the latter in \eqref{7-12} we write an equation that describes a coupled graviton (torsion) with a scalar particle
\begin{eqnarray}
\Phi^{(1)\mu\nu}= \epsilon^{\mu\alpha\beta} \partial_\alpha {{h^{Tt}}^\nu}_\beta + m {h^{Tt}}^{\mu\nu } 
+m(\eta^{\mu\nu}+ \hat{\partial}_\mu\hat{\partial}_\nu) \phi-\phi h^{\mu\nu}(h^{Tt},\phi) + \nonumber \\+
{h^\mu}_\lambda(h^{Tt},\phi) k^{\nu\lambda}(h^{Tt},\phi)\,\,, \label{7-18}
\end{eqnarray}
and the trace of this is
\begin{eqnarray}
{\Phi^{(1)\mu}}_\mu = 4m\phi-\frac{3}{4}\,\phi^2 +{h^{Tt}}_{\alpha\beta}{h^{Tt}}^{\alpha\beta}
+2{h^{Tt}}_{\alpha\beta}\hat{\partial}^\alpha\hat{\partial}^\beta\phi\,\,, \label{7-19}
\end{eqnarray}
and together with \eqref{7-18} show that if one ''turns off'' the quadratic terms (weak Weitzenb\"{o}ck regime), one will be
describing a physical system with single and unitary selfdual spin 2 propagation. Thus, we see that the effect of the cubic potential of self-interaction
is to excite the scalar propagation in such a way that the number of degrees of freedom coincides with that of the Mielke-Baekler model, as it could be expected. To finish this review, we want to confirm that the unitarity of the physical system is preserved in the passage from a pure selfdual spin $2$ theory to a modified type given by \eqref{7-11}.

\subsection{Non-covariant decomposition and energy}

Looking at \eqref{7-11b} one can think that the computation of the exchange amplitude density ($\mathcal{A}$) for this system it is a cumbersome task. In fact it is. When an external source is minimally coupled with the field $k_{\mu\nu}$ the non-linear field equations give rise to non-linear relation for fields in terms of the external source. However, in the same way that it could be performed the reduction of field equations which shows the existence of two degrees of freedom (a massive graviton and a massive scalar), the construction of a reduced action coming from \eqref{7-11b} seems to be an available procedure. A short path to verify unitarity is the following. let it start with a non-covariant TLt decomposition for field $k_{\mu\nu}$ 
\begin{eqnarray}
k_{00}\equiv n\,, \label{7-20}
\end{eqnarray}
\begin{eqnarray}
k_{0i}\equiv N_i\,, \label{7-21}
\end{eqnarray}
\begin{eqnarray}
k_{i0}\equiv M_i\,, \label{7-22}
\end{eqnarray}
\begin{eqnarray}
k_{ij}\equiv h_{ij}+\epsilon_{ij}v\,, \label{7-23}
\end{eqnarray}
where $h_{ij}\equiv k_{\{ij\}}$ and $v\equiv\frac{\epsilon_{ij}}2\, k_{ij}$. Then, the action  \eqref{7-11b}
looks like
\begin{eqnarray}
S=\langle  -\epsilon_{ij} N_i \dot{N}_j +(\epsilon_{ij}h_{ni}-2\eta_{nj}v)\dot{h}_{nj} - m (h_{ij}h_{ij}-{h^2}_{ii} - 2v^2) + 
\nonumber \\+n[2\epsilon_{ij} \partial_i N_j-2mh_{ii}+ {h^2}_{ii} -h_{ij}h_{ij} + 2v^2] +
\nonumber \\+M_r[ -2\epsilon_{ij} \partial_i(h_{rj}+\epsilon_{rj}v)+2(m -h_{ii})N_r+2N_i(h_{ir}+\epsilon_{ir}v)] \rangle\,. \label{7-24}
\end{eqnarray}
It can be noted at a first sight that fields $n$ and $M_{n}$ do appear as Lagrange's multipliers and do not contribute to dynamics. However,  
in order to constuct a two degree of freedom reduced action there is more available
constraints. Constraints \eqref{7-15} and \eqref{7-16} are now
\begin{eqnarray}
N_i=M_i\,, \label{7-25}
\end{eqnarray}
\begin{eqnarray}
v=0\,, \label{7-26}
\end{eqnarray}
\begin{eqnarray}
\dot{N}_i +\partial_i h_{jj}-\partial_j h_{ij}-\partial_in=0\,, \label{7-27}
\end{eqnarray}
\begin{eqnarray}
\dot{h}_{ii}-\partial_iN_i=0\,, \label{7-28}
\end{eqnarray}
which allow us to write down a first approach to reduced action coming from \eqref{7-24}, in other words
\begin{eqnarray}
S^* =\langle  \epsilon_{ij}h_{ni}\dot{h}_{nj} - m (h_{ij}h_{ij}-{h^2}_{ii}) +n[\epsilon_{ij} \partial_i N_j-2mh_{ii}+ {h^2}_{ii} -h_{ij}h_{ij}] +
\nonumber \\+N_r[ -2\epsilon_{ij} \partial_ih_{rj}+2(m -h_{ii})N_r+2N_ih_{ir}-\epsilon_{ri}(\partial_jh_{ji}-\partial_ih_{jj})] \rangle \,.\nonumber \\
 \label{7-29}
\end{eqnarray}
This relation tell us that the physical content is enclosed in the traceless field ${h^t}_{ij}$ because, on one hand the constraint related to the multiplier 
$n$ (i.e., $\epsilon_{ij} \partial_i N_j-2mh_{ii}+ {h^2}_{ii} -h_{ij}h_{ij}=0$) means a relation for the trace $h_{ii}$ and, on the other hand, 
the constraint associated to the multiplier $N_r$ (i.e., $-2\epsilon_{ij} \partial_ih_{rj}+4(m -h_{ii})N_r+4N_ih_{ir}-
\epsilon_{ri}(\partial_jh_{ji}-\partial_ih_{jj})=0$) allow to solve this field in terms of $h_{ij}$. In this sense, one can introduce a non-covariant 
traceless decomposition given by
\begin{eqnarray}
h_{ij}={h^t}_{ij}+\frac{\eta_{ij}}2\,h_{ll}\,, \label{7-30}
\end{eqnarray}
where ${h^t}_{ll}=0$. Now, the action \eqref{7-24} can be written again as 
\begin{eqnarray}
S^* =\langle  \epsilon_{ij}{h^t}_{ni}\dot{h^t}_{nj} - m ({h^t}_{ij}{h^t}_{ij}-\frac{{h^2}_{ii}}2) +n[\epsilon_{ij} \partial_i N_j-2mh_{ii}
+ \frac{{h^2}_{ii}}2 
-{h^t}_{ij}{h^t}_{ij}] +
\nonumber \\+N_r[ -2\epsilon_{ij} \partial_i{h^t}_{rj}-\epsilon_{ri}\partial_j{h^t}_{ji}+2N_i{h^t}_{ir}+(2m -h_{ii})N_r-\frac{3}2\,
\epsilon_{ir}\partial_ih_{jj})] \rangle\,,\nonumber \\
 \label{7-31}
\end{eqnarray}
then using the constraints related to $n$ and $N_r$ and dropping out the non-dynamical $h_{ll}$, one can say that the reduced action is
\begin{eqnarray}
S^* =\langle  \epsilon_{ij}{h^t}_{ni}\dot{h^t}_{nj} - m {h^t}_{ij}{h^t}_{ij} \rangle\,,
 \label{7-31b}
\end{eqnarray}
which explicitely shows the correct sign  at the massive term (i.e., positively defined energy) and two degree of freedom.

\section{Topologically massive gravity in the Einstein-Weyl space-time}

In contrast with the last section, now we impose the condition on torsion field given by ${T^\lambda}_{\mu\nu}=0$ (so, ${K^\lambda}_{\mu\nu}=0$), meanwhile metric and non-metricity are dynamical fields. From \eqref{ecu:15} we write 
\begin{equation}
A^{\lambda}{_{\mu\nu}} = {\Gamma^\lambda}_{\mu\nu}(g) - {\gamma^\lambda}_{\mu\nu}(Q) \,\,\,\,, \label{8-1}
\end{equation}
or in a matricial way $ A_\mu = \Gamma_\mu - \gamma_\mu $. We remember that $ \gamma_\mu (Q)$ are the non-metricity symbols whose components are defined in \eqref{ecu:18} (i.e., ${\gamma^\lambda}_{\mu \nu}(Q) \equiv \dfrac{g^{\lambda \alpha}}{2}\,(Q_{\mu\alpha\nu} + Q_{\nu\alpha\mu} - Q_{\alpha\mu\nu} ) $). Due to the obvious symmetry property, ${\gamma^\lambda}_{\mu \nu}={\gamma^\lambda}_{\nu \mu}$ it can expected more degrees of freedom than in the Riemann-Cartan case arising spins $3$, $2$, $1$ and $0$.

Next, the Riemann tensor (matricially) is $\mathbf{R}_{\mu\nu} = \partial_\nu A_\mu - \partial_\mu A_\nu + \left[ A_\nu, A_\mu \right]$ and in terms of the Levi-Civita derivative is 
\begin{equation}
\mathbf{R}_{\mu\nu} = \mathbf{R}_{\mu\nu}(\Gamma) - \left[\gamma_\mu, \gamma_\nu \right] +\nabla_\mu \gamma_\nu-\nabla_\nu \gamma_\mu \,, \label{8-2}
\end{equation}
and the Ricci-Weyl tensor and scalar are
\begin{equation}
R_{\beta\nu} = R_{\beta\nu}(\Gamma) - {\left[ \gamma_\alpha, \gamma_\nu \right]^\alpha}_\beta+ \nabla_\alpha {\gamma^\alpha}_{\nu\beta}- \nabla_\nu {\gamma^\alpha}_{\alpha\beta} \,, \label{8-3}
\end{equation}
\begin{equation}
R = R(\Gamma) - g^{\nu\beta}{\left[ \gamma_\alpha,\gamma_\nu \right]^\alpha}_\beta + \nabla_\sigma (g^{\alpha\beta}{\gamma^\sigma}_{\alpha\beta} -g^{\sigma\beta}{\gamma^\alpha}_{\alpha\beta}) \,, \label{8-4}
\end{equation}
respectively. Then the Hilbert-Einstein-Weyl action is
\begin{equation}
S_{HE}[g,\gamma] = -\sigma \langle R(\Gamma) - g^{\nu\beta}{\left[ \gamma_\alpha,\gamma_\nu \right]^\alpha}_\beta  \rangle \,. \label{8-5}
\end{equation}

In Einstein-Weyl space the Chern-Simons action is
\begin{eqnarray}
S_{CS}[g,\gamma] = S_{CS}(\Gamma) - \dfrac{\sigma}{2m} \langle \varepsilon^{\mu\nu\lambda} tr\,( \gamma_\mu \nabla_\nu \gamma_\lambda + \dfrac{2}{3}\gamma_\mu \gamma_\nu \gamma_\lambda )  \rangle\,,  \label{8-6}
\end{eqnarray}
and the action for TMG in Einstein-Weyl (TMGEW) is
\begin{eqnarray}
S_{TMGEW} = S_{TMG} [\Gamma] + \tilde{S}_{CS} [g,\gamma] \,, \label{8-7}
\end{eqnarray}
where again $S_{TMG}[\Gamma]$ is the action for 
TMG in  Riemann-Christoffel, eq. \eqref{7-9} and $\tilde{S}_{CS} [g,\gamma]$ is a modified tensor-Chern-Simons given by
\begin{eqnarray}
\tilde{S}_{CS}[g,\gamma] = \dfrac{\sigma}{m}\langle  -\frac{\varepsilon^{\mu\nu\lambda}}2\,tr( \gamma_\mu \nabla_\nu \gamma_\lambda +\frac{2}3\,\gamma_\mu \gamma_\nu \gamma_\lambda) + m  g^{\nu\beta}{\left[ \gamma_\alpha,\gamma_\nu \right]^\alpha}_\beta\rangle \,, \label{8-8}
\end{eqnarray}
who we shall study next in a similar way we had explored the Riemann-Cartan context. The main objective is to reveal the unitarity problems residing in this model and it is enough to consider a weak Einstein-Weyl regime as a first step. In this sense let us to define the second order in field power action at the limit $g_{\mu\nu}\rightarrow \eta_{\mu\nu}$ as follows
\begin{eqnarray}
S[\eta,\gamma] = \langle  -\frac{\epsilon^{\mu\nu\lambda}}2\,tr \gamma_\mu \partial_\nu \gamma_\lambda + m  \eta^{\nu\beta}{\left[ \gamma_\alpha,\gamma_\nu \right]^\alpha}_\beta\rangle \,. \label{8-9}
\end{eqnarray}

Now we need to stablish some decomposition for the non-metricity symbols starting with a semi-traceless one
\begin{eqnarray}
\gamma_{\lambda\mu\nu}= \omega_{\lambda\mu\nu} +2\eta_{\lambda \{\mu}c_{\nu\}}  \,, \label{8-10}
\end{eqnarray}
where $\omega_{\lambda\mu\nu}=\omega_{\lambda\nu\mu}$ and 
\begin{eqnarray}
{{\omega_\lambda}^\mu}_\mu=0 \,, \label{8-11}
\end{eqnarray}

Using  \eqref{8-10} in  \eqref{8-9} give rise
\begin{eqnarray}
S[\omega,c] = \langle  -\frac{\epsilon^{\mu\nu\lambda}}2\,{\omega^\alpha}_{\mu\beta} \partial_\nu {\omega^\beta}_{\lambda\alpha}-m{\omega^\alpha}_{\mu\beta}{\omega^{\beta\mu}}_\alpha-2\epsilon^{\mu\nu\lambda}c_\mu\partial_\nu c_\lambda
+2 mc_\mu c^\mu +\nonumber \\-\,  \epsilon^{\mu\nu\lambda}{\omega^\sigma}_{\sigma\mu}\partial_\nu c_\lambda+\lambda^\alpha {{\omega_\alpha}^\mu}_\mu\rangle \,\,,\,\,\,\label{8-12}
\end{eqnarray}
where we have explicitely added the constraint \eqref{8-11} through the multiplier $\lambda^\alpha$. From \eqref{8-12} it  can be clearly seen one of many problems in TMGEW, this means the wrong sign at the Proca term ''$2 mc_\mu c^\mu $'' which shall carries non-unitary excitation. The field equations are 
\begin{eqnarray}
0={\Phi^{\beta\mu}}_\alpha \equiv -\epsilon^{\{\mu\rho\sigma} \partial_\rho {\omega^{\beta\}}}_{\sigma\alpha} 
+\frac{\eta^{\mu\beta}}3\,\epsilon^{\rho\sigma\lambda} \partial_\sigma \omega_{\rho\lambda\alpha} 
-2m({\omega^{\{\mu\beta\}}}_\alpha  -\frac{\eta^{\mu\beta}}3\, {\omega^\rho}_{\rho\alpha})+\nonumber \\+\,\frac{3}5\,{\delta^{\{\mu}}_\alpha (\frac{\epsilon^{\beta\}\rho\sigma}}2\,\partial_\rho {\omega^\lambda}_{\lambda\sigma}+\frac{\epsilon^{\sigma\rho\lambda}}3\,\partial_\sigma {\omega_{\rho\lambda}}^{\beta\}}+\frac{m}3\,{{\omega^\lambda}_\lambda}^{\beta\}})\nonumber \\-\,
\frac{\eta^{\mu\beta}}5\, (\frac{{\epsilon_\alpha}^{\rho\sigma}}2\,\partial_\rho {\omega^\lambda}_{\lambda\sigma}+\frac{\epsilon^{\sigma\rho\lambda}}3\,\partial_\sigma \omega_{\rho\lambda\alpha}+\frac{m}3\,{\omega^\lambda}_{\lambda\alpha})
\,\,\,,\nonumber \\ \label{8-13}
\end{eqnarray}
\begin{eqnarray}
0=\Psi^\mu \equiv 4m {c^T}^\mu+4\frac{\epsilon^{\sigma\rho\lambda}}5\,\partial_\sigma {\omega_{\rho\lambda}}^\mu +\frac{\epsilon^{\mu\rho\sigma}}5\,\partial_\rho {\omega^\lambda}_{\lambda\sigma} +\frac{4m}5\, {{\omega^\lambda}_\lambda}^\mu \,\,\,,\label{8-14}
\end{eqnarray}
and \eqref{8-11}, of course. It can be observed that the longitudinal part of $c_\mu$ (i.e., $c^L$) is naturally dropped out from theory staying only the transverse one, ${c^T}_\mu$. This transversality of $c_\mu$ remains at full theory with cubic terms provided. Additionally, divergence of \eqref{8-14} gives rise to a scalar constraint over spin $0$ fields, in other words
\begin{eqnarray}
0=\frac{5}4\,\partial_\mu\Psi^\mu \equiv \epsilon^{\sigma\rho\lambda}\partial_\mu\partial_\sigma {\omega_{\rho\lambda}}^\mu +\partial_\mu{{\omega^\lambda}_\lambda}^\mu \,\,\,.\label{8-15}
\end{eqnarray}
Both the last equation and \eqref{8-14} say that from a set of four spin $0$ (three in $\omega_{\alpha\mu\nu}$ and one in $c_\mu$) only survive two of them  (unfortunately they are thaquions, as we shall see). Next we explore these aspects, among other ones using the illustrative procedure of the TLt field decomposition (see Appendix B) which allows to discern the $3$, $2$, $1$ and $0$ spin sectors. Let us recall the TLt decomposition for $\omega_{\alpha\mu\nu}=\omega_{\alpha\nu\mu}$ with ${{\omega_\alpha}^\mu}_\mu=0$, this means
\begin{eqnarray}
\omega_{\alpha\mu\nu}={S^{Tt}}_{\alpha\mu\nu}+\hat\partial_{(\alpha}{s^{Tt}}_{\mu\nu)}+2\hat\partial_{\{\mu}{\tau^{Tt}}_{\alpha\nu\}}
+[\hat\partial_{(\alpha}\hat\partial_\mu -\frac{\eta_{(\alpha\mu}}5]{w^T}_{\nu)}+2\hat\partial_{\{\mu}\hat\partial_\alpha{b^T}_{\nu\}}+
\nonumber \\  +2 {\epsilon_{\alpha\{\mu}}^\lambda(\hat\partial_{\nu\}}{v^T}_\lambda+\hat\partial_\lambda{v^T}_{\nu\}})+
[\hat\partial_\alpha\hat\partial_\mu\hat\partial_\nu -\frac{\eta_{(\alpha\mu}}5\,\hat\partial_{\nu)}]w+2 {\epsilon_{\alpha\{\mu}}^\lambda\hat\partial_{\nu\}}\hat\partial_\lambda\theta+\nonumber \\ -2(\hat\partial_\alpha\hat\partial_\mu\hat\partial_\nu -\eta_{\alpha\{\mu}\hat\partial_{\nu\}})\psi
 \,\,\,,   \nonumber \\
\label{8-16}
\end{eqnarray}
and 
\begin{eqnarray}
c_\mu= {c^T}_\mu +\hat\partial_\mu{c^L}\,\,\,,\label{8-17}
\end{eqnarray}
where the symbol ''()'' means a cyclic sum (i.e., $(\alpha\mu\nu)\equiv \alpha\mu\nu +\mu\nu\alpha+ \nu\alpha\mu$), ''$Tt$'' for transverse-traceless, etc. 

Inserting this decomposition in \eqref{8-12} we can rewrite the action as a sum of spin $3$, $2$, $1$ and $0$ sectors, respectively 
\begin{eqnarray}
S[\omega,c] = S^{(3)}[S^{Tt}] +S^{(2)}[s^{Tt},\tau^{Tt}]+S^{(1)}[w^T,b^T,v^T,c^T]+S^{(0)}[w,\theta,\psi,c^L]\,\,\,,\nonumber \\
\label{8-18}
\end{eqnarray}
where
\begin{eqnarray}
S^{(3)}\equiv\langle  -\frac{\epsilon^{\mu\nu\lambda}}2\,{{S^{Tt}}^\alpha}_{\mu\beta} \partial_\nu {{S^{Tt}}^\beta}_{\lambda\alpha}
-m{S^{Tt}}_{\alpha\mu\beta}{S^{Tt}}^{\alpha\mu\beta}\rangle \,\,\,,\label{8-19}
\end{eqnarray}
\begin{eqnarray}
S^{(2)}\equiv\langle  \epsilon^{\mu\nu\lambda}{s^{Tt}}_{\mu\beta} \partial_\nu( {{s^{Tt}}_\lambda}^\beta + {{\tau^{Tt}}_\lambda}^\beta) +
3m{s^{Tt}}_{\alpha\mu}{s^{Tt}}^{\alpha\mu}+4m{s^{Tt}}_{\alpha\mu}{\tau^{Tt}}^{\alpha\mu}+\nonumber \\
+\,m{\tau^{Tt}}_{\alpha\mu}{\tau^{Tt}}^{\alpha\mu}\rangle \,\,\,,\label{8-20}
\end{eqnarray}
\begin{eqnarray}
S^{(1)}\equiv\langle  -\frac{2}5\,\epsilon^{\mu\nu\lambda}{w^T}_\mu \partial_\nu {w^T}_\lambda 
-\frac{\epsilon^{\mu\nu\lambda}}2\,{b^T}_\mu \partial_\nu {b^T}_\lambda +
2\epsilon^{\mu\nu\lambda}{v^T}_\mu \partial_\nu {v^T}_\lambda\nonumber + \\-
\frac{4}5\,\epsilon^{\mu\nu\lambda}{w^T}_\mu \partial_\nu {b^T}_\lambda
-2\epsilon^{\mu\nu\lambda}{c^T}_\mu \partial_\nu {c^T}_\lambda -\epsilon^{\mu\nu\lambda}{b^T}_\mu \partial_\nu {c^T}_\lambda +\nonumber \\
+\frac{8}5\,{w^T}_\mu \Box^{\frac{1}2} {v^T}^\mu +{b^T}_\mu \Box^{\frac{1}2} {v^T}^\mu -\frac{12m}5\,{w^T}_\mu {w^T}^\mu -m{b^T}_\mu {b^T}^\mu +\nonumber \\
+6m{v^T}_\mu {v^T}^\mu+2m{c^T}_\mu {c^T}^\mu-\frac{16m}5\,{w^T}_\mu {b^T}^\mu-2m\epsilon^{\mu\nu\lambda}{b^T}_\mu \hat\partial_\nu {v^T}_\lambda
\rangle \,\,\,,\label{8-21}
\end{eqnarray}
\begin{eqnarray}
S^{(0)}\equiv\langle 
-\frac{2}5\,w \Box^{\frac{1}2} \theta +\frac{2m}5\,w^2-2m\theta^2+2m\psi^2-\frac{8m}5\,w\psi-2m{c^L}^2
\rangle \,\,\,.\label{8-22}
\end{eqnarray}
Next we discuss the physical content of every spin sector.

\subsection{Spin 3.}

Action \eqref{8-19} represents a selfdual spin 3 propagation with a first order equation given by
\begin{eqnarray}
{\epsilon_\alpha}^{\sigma \rho}\partial_\sigma {S^{Tt}}_{\rho\mu\nu}+2m{S^{Tt}}_{\alpha\mu\nu}=0\,\,\,,\label{8-23}
\end{eqnarray}
or in a second order form
\begin{eqnarray}
[\Box-(2m)^2]{S^{Tt}}_{\alpha\mu\nu}=0\,\,\,,\label{8-24}
\end{eqnarray}
which represents a causal degree of freedom with mass equal to $2m$. The correct sign of the massive term in \eqref{8-19} obviously tell us about unitarity of this propagation. However, in order to compare to other spin sectors, we review the density of one particle exchange amplitude as follows. Let $S^{(3)}[j]$ be action including interaction with an external source 
\begin{eqnarray}
S^{(3)}[j]=\langle  -\frac{\epsilon^{\mu\nu\lambda}}2\,{{S^{Tt}}^\alpha}_{\mu\beta} \partial_\nu {{S^{Tt}}^\beta}_{\lambda\alpha}
-m{S^{Tt}}_{\alpha\mu\beta}{S^{Tt}}^{\alpha\mu\beta}-k j_{\alpha\mu\beta}{S^{Tt}}^{\alpha\mu\beta}\rangle \,\,\,,\label{8-25}
\end{eqnarray}
giving rise the field equation
\begin{eqnarray}
{\epsilon_\alpha}^{\sigma \rho}\partial_\sigma {S^{Tt}}_{\rho\mu\nu}+2m{S^{Tt}}_{\alpha\mu\nu}+k j_{\alpha\mu\nu}=0\,\,\,.\label{8-26}
\end{eqnarray}
Consistency with degree of freedom count demands that $j_{\alpha\mu\nu}$ must be a transverse (conserved), traceless and totally symmetric field. Using the second orden equation coming from \eqref{8-26}, the field ${S^{Tt}}_{\rho\mu\nu}$ can be formally solved in terms of the external source as follows
\begin{eqnarray}
{S^{Tt}}_{\rho\mu\nu}[j]= 2mkD(2m)j_{\rho\mu\nu}-k{\epsilon_\rho}^{\alpha \beta}\partial_\alpha D(2m)j_{\beta\mu\nu}\,\,\,,\label{8-27}
\end{eqnarray}
where $D(2m)$ is the (dual) propagator
\begin{eqnarray}
D(2m)=\frac{1}{\Box-(2m)^2}\,\,\,.\label{8-28}
\end{eqnarray}
Then, the one particle exchanje amplitude density between two sources is 
\begin{eqnarray}
\mathcal{A}\equiv {j'}^{\rho\mu\nu} {S^{Tt}}_{\rho\mu\nu}[j]= 2mk\,{j'}^{\rho\mu\nu}D(2m)j_{\rho\mu\nu}-k{j'}^{\rho\mu\nu}{\epsilon_\rho}^{\alpha \beta}\partial_\alpha D(2m)j_{\beta\mu\nu}\,\,\,,\label{8-29}
\end{eqnarray}
but in this way does not say much. Following the well known idea about pseudospin projectors \cite{Aragone} one can rewrite the exchange density $\mathcal{A}$ in order to show clearly unitarity or non-unitarity. Let us to introduce the following object 
\begin{eqnarray}
{{\Omega^{Tt}}_{\alpha\mu\nu}}^{(s)}= p{\Omega^{Tt}}_{\alpha\mu\nu}+sr{\epsilon_\alpha}^{\sigma \rho}\hat\partial_\sigma{\Omega^{Tt}}_{\rho\mu\nu}\,\,\,,\label{8-30}
\end{eqnarray}
for some ${\Omega^{Tt}}_{\alpha\mu\nu}$ where $s=\pm 1$, $p$ a positive-definite real number and $r$ a real number. Taking $p=r=\frac{1}{\sqrt{2}}$ and  ${\Omega^{Tt}}_{\alpha\mu\nu}=j_{\alpha\mu\nu}$ it can be written the next expression
\begin{eqnarray}
{{j'}^{\rho\mu\nu}}^{(s)}D(2m){j_{\rho\mu\nu}}^{(s)}= {j'}^{\rho\mu\nu}D(2m)j_{\rho\mu\nu}+s{j'}^{\rho\mu\nu}{\epsilon_\rho}^{\alpha \beta}\hat\partial_\alpha D(2m)j_{\beta\mu\nu} \,\,\,,\label{8-31}
\end{eqnarray}
up to a divergence. In order to compare \eqref{8-31} with \eqref{8-29} we use the following identity 
\begin{eqnarray}
D(2m)\,\frac{\partial_\mu}{2m}\equiv D'(2m)\hat\partial_\mu \,\,\,,\label{8-32}
\end{eqnarray}
with $D'(2m)=D(2m)+1/[2m(\Box^{\frac{1}2}+2m)]$, which state that $D'(2m)$ reveals the same pole at $\Box=(2m)^2$. So, from \eqref{8-29} we say
\begin{eqnarray}
\frac{\mathcal{A}}{2mk}=  {j'}^{\rho\mu\nu}D(2m)j_{\rho\mu\nu}-{j'}^{\rho\mu\nu}{\epsilon_\rho}^{\alpha \beta}\hat\partial_\alpha D(2m)j_{\beta\mu\nu}=\nonumber \\ = {{j'}^{\rho\mu\nu}}^{(-1)}D(2m){j_{\rho\mu\nu}}^{(-1)}\,\,\,,\label{8-33}
\end{eqnarray}
up to an holomorphic function at the pole $\Box=(2m)^2$. The residue is $Res(\frac{\mathcal{A}}{2mk})=+1$ confirming unitarity. An analogous procedure shall be applied in the next spin sectors.

\subsection{Spin 2.}

Just for convenience let it try a variable change given by ${\Phi^{Tt}}_{\mu\beta}={s^{Tt}}_{\mu\beta}+{\tau^{Tt}}_{\mu\beta}$ with which the action \eqref{8-20} can be reformulated as $S^{(2)}=\langle  \epsilon^{\mu\nu\lambda}{s^{Tt}}_{\mu\beta} \partial_\nu{{\Phi^{Tt}}_\lambda}^\beta +
2m{s^{Tt}}_{\alpha\mu}{\Phi^{Tt}}^{\alpha\mu}+m{\Phi^{Tt}}_{\alpha\mu}{\Phi^{Tt}}^{\alpha\mu}\rangle $, although anyway the signs at the mass terms are suspects of carryng non-unitarity. Then, the theory with interaction is described trough the following action
\begin{eqnarray}
S^{(2)}[j_1,j_2]=\langle  \epsilon^{\mu\nu\lambda}{s^{Tt}}_{\mu\beta} \partial_\nu{{\Phi^{Tt}}_\lambda}^\beta +
2m{s^{Tt}}_{\alpha\mu}{\Phi^{Tt}}^{\alpha\mu}+m{\Phi^{Tt}}_{\alpha\mu}{\Phi^{Tt}}^{\alpha\mu}+ \nonumber \\
-k{s^{Tt}}_{\alpha\mu}{j_1}^{\alpha\mu}-k{\Phi^{Tt}}_{\alpha\mu}{j_2}^{\alpha\mu}\rangle \,\,\,,\label{8-34}
\end{eqnarray}
and field equations are
\begin{eqnarray}
{\epsilon_\mu}^{\sigma\rho}\partial_\sigma{\Phi^{Tt}}_{\rho\nu}  +
2m{\Phi^{Tt}}_ {\mu\nu}=k{j_1}_{\mu\nu} \,\,\,,\label{8-35}
\end{eqnarray}
\begin{eqnarray}
{\epsilon_\mu}^{\sigma\rho}\partial_\sigma{s^{Tt}}_{\rho\nu} +2m{s^{Tt}}_ {\mu\nu} +
2m{\Phi^{Tt}}_ {\mu\nu}=k{j_2}_{\mu\nu} \,\,\,,\label{8-36}
\end{eqnarray}
where current sources must be symmetric, transverse and traceless for a consistent coupling.

Using \eqref{8-35} and \eqref{8-36}, the second order field equation are 
\begin{eqnarray}
[\Box-(2m)^2]{\Phi^{Tt}}_ {\alpha\nu}=-2mk{j_1}_{\alpha\nu}+k {\epsilon_\alpha}^{\sigma\rho}\partial_\sigma{j_1}_{\rho\nu}  \,\,\,,\label{8-37}
\end{eqnarray}
\begin{align}
[\Box-(2m)^2]{s^{Tt}}_{\rho\nu}&=-2mk{j_2}_{\alpha\nu}+k {\epsilon_\alpha}^{\sigma\rho}\partial_\sigma{j_2}_{\rho\nu} +\nonumber \\ 
&-2mk[\Box+(2m)^2]D(2m){j_1}_{\alpha\nu}+2(2m)^2k {\epsilon_\alpha}^{\sigma\rho}\partial_\sigma D(2m){j_1}_{\rho\nu}\,\,\,,\label{8-38}
\end{align}
and fields are solved formally
\begin{eqnarray}
{\Phi^{Tt}}_ {\alpha\nu}[j_1]=-2mkD(2m){j_1}_{\alpha\nu}+k {\epsilon_\alpha}^{\sigma\rho}\partial_\sigma D(2m){j_1}_{\rho\nu}  \,\,\,,\label{8-39}
\end{eqnarray}
\begin{align}
{s^{Tt}}_{\rho\nu}[j_1,j_2]&=-2mkD(2m){j_2}_{\alpha\nu}+k {\epsilon_\alpha}^{\sigma\rho}\partial_\sigma D(2m){j_2}_{\rho\nu} +\nonumber \\ 
&-2mk[\Box+(2m)^2]D^2(2m){j_1}_{\alpha\nu}+2(2m)^2k {\epsilon_\alpha}^{\sigma\rho}\partial_\sigma D^2(2m){j_1}_{\rho\nu}\,\,\,.\label{8-40}
\end{align}
Putting together \eqref{8-27} and \eqref{8-39} one can see in this last case that opossite sign conduces to a non-unitary propagation (graviton ghost).
The situation for ${s^{Tt}}_{\rho\nu}$ in \eqref{8-40} gets worst because the quadratical power of $D(2m)$ which conduces to an  ill defined exchange amplitude.

\subsection{Spin 1.}

Viewing \eqref{8-21}, the inclusion of interaction conduces to the next action
\begin{eqnarray}
S^{(1)}[j_1,j_2,j_3,j_4]=\langle  -\frac{2}5\,\epsilon^{\mu\nu\lambda}{w^T}_\mu \partial_\nu {w^T}_\lambda 
-\frac{\epsilon^{\mu\nu\lambda}}2\,{b^T}_\mu \partial_\nu {b^T}_\lambda +
2\epsilon^{\mu\nu\lambda}{v^T}_\mu \partial_\nu {v^T}_\lambda\nonumber + \\-
\frac{4}5\,\epsilon^{\mu\nu\lambda}{w^T}_\mu \partial_\nu {b^T}_\lambda
-2\epsilon^{\mu\nu\lambda}{c^T}_\mu \partial_\nu {c^T}_\lambda -\epsilon^{\mu\nu\lambda}{b^T}_\mu \partial_\nu {c^T}_\lambda +\nonumber \\
+\frac{8}5\,{w^T}_\mu \Box^{\frac{1}2} {v^T}^\mu +{b^T}_\mu \Box^{\frac{1}2} {v^T}^\mu -\frac{12m}5\,{w^T}_\mu {w^T}^\mu -m{b^T}_\mu {b^T}^\mu +\nonumber \\
+6m{v^T}_\mu {v^T}^\mu+2m{c^T}_\mu {c^T}^\mu-\frac{16m}5\,{w^T}_\mu {b^T}^\mu-2m\epsilon^{\mu\nu\lambda}{b^T}_\mu \hat\partial_\nu {v^T}_\lambda +\nonumber \\
-k{w^T}_\mu{j_1}^\mu -k{b^T}_\mu{j_2}^\mu -k{v^T}_\mu{j_3}^\mu -k{c^T}_\mu{j_4}^\mu \rangle \,\,\,.\nonumber \\ \label{8-41}
\end{eqnarray}

In order to handle the first order field equations which occur from the last action, let us recall some notation about projectors (appendix B). Let ${\Theta_\mu}^\nu\equiv {\epsilon_\mu}^{\sigma\nu}\hat\partial_\sigma$ a parity operator with the property ${\Theta_\mu}^\nu{\Theta_\nu}^\lambda={P_\mu}^\lambda$, where ${P_\mu}^\lambda= {\delta_\mu}^\lambda-\hat\partial_\mu\hat\partial^\lambda$ is a transverse projector. Then, given a transverse object, ${\Omega^T}_\mu$, we shall call the ''dual'' of this one to
\begin{align}
{\bar\Omega^T}\,_\mu\equiv {\Theta_\mu}^\nu {\Omega^T}_\nu\,\,\,,\label{8-42}
\end{align}
and the inverse
\begin{align}
{\Omega^T}_\mu\equiv {\Theta_\mu}^\nu {\bar\Omega^T}\,_\nu\,\,\,,\label{8-43}
\end{align}
this means ${\bar{\bar\Omega}^T}_\mu={\Omega^T}_\mu$. Then, introducing the ''mass operator'', $\mu\equiv m\Box^{-\frac{1}2}$ and the $hat-current$, $\hat{j}_{a\mu}\equiv \Box^{-\frac{1}2}j_{a\mu}$ with $a=1,2,3,4$, the field equations coming from \eqref{8-41} are 
\begin{eqnarray}
{\bar w^T}_\lambda 
+{\bar b^T}_\lambda 
+6\mu{w^T}_ \lambda +4\mu {b^T}_\lambda -2 {v^T}_\lambda   =-\frac{5}4\,k\hat{j}_{1\lambda} \,\,\,, \label{8-44}
\end{eqnarray}
\begin{eqnarray}
\frac{4}5\, {\bar w^T}_\lambda 
+ {\bar b^T}_\lambda +2\mu{\bar v^T}_\lambda + {\bar c^T}_\lambda
 +\frac{16\mu}5\,{w^T}_\lambda+2\mu{b^T}_\lambda-{v^T}_\lambda =-k\hat{j}_{2\lambda} \,\,\,, \label{8-45}
\end{eqnarray}
\begin{eqnarray}
-\mu{\bar b^T}_\lambda +2 {\bar v^T}_\lambda 
 +\frac{4}5\, {w^T}_\lambda
+\frac{1}2\,{b^T}_\lambda +6\mu{v^T}_\lambda =\frac{k}2\,\hat{j}_{3\lambda}  \,\,\,, \label{8-46}
\end{eqnarray}
\begin{eqnarray}
{\bar b^T}_\lambda+4{\bar c^T}_\lambda 
 -4\mu{c^T}_\lambda =-k\hat{j}_{4\lambda} \,\,\,, \label{8-47}
\end{eqnarray}
where all currents are transverse. From this system, for example all fields can be written in terms of the field ${c^T}_\lambda$ and currents, in other words
\begin{eqnarray}
{w^T}_\lambda=-\frac{5\mu}3\, {\bar c^T}_\lambda +\frac{5}3\,(2+\mu^2) {c^T}_\lambda +kl_{1 \lambda} \,\,\,, \label{8-48}
\end{eqnarray}
\begin{eqnarray}
{b^T}_\lambda=4\mu {\bar c^T}_\lambda -4 {c^T}_\lambda +kl_{2 \lambda} \,\,\,, \label{8-49}
\end{eqnarray}
\begin{eqnarray}
{v^T}_\lambda=\frac{1}3\,(-1+4\mu^2) {\bar c^T}_\lambda +\frac{2\mu}3\,{c^T}_\lambda +kl_{3 \lambda} \,\,\,, \label{8-50}
\end{eqnarray}
where
\begin{eqnarray}
l_{1 \lambda}\equiv -\frac{25}{24\mu}\,\hat{j}_{1\lambda}-\frac{25}{18}\,\bar{\hat{j}}_{1\lambda} +\frac{5}{4\mu}\,\hat{j}_{2\lambda} +\frac{25}{12}\,\bar{\hat{j}}_{2\lambda}+\frac{5}{8}(\frac{5}9-\frac{1}{2\mu})\,\hat{j}_{3\lambda}+\frac{25}{76\mu}\,\bar{\hat{j}}_{3\lambda}+\nonumber \\
-(\frac{95}{288\mu}+\frac{5\mu}{12})\,\hat{j}_{4\lambda}\,\,\,, \label{8-51}
\end{eqnarray}
\begin{eqnarray}
l_{2 \lambda}\equiv -\bar{\hat{j}}_{4\lambda}\,\,, \label{8-52}
\end{eqnarray}
\begin{eqnarray}
l_{3 \lambda}\equiv -\frac{10}{9}\,\bar{\hat{j}}_{1\lambda} +\frac{5}{3}\,\bar{\hat{j}}_{2\lambda}+\frac{5}{18}\,\hat{j}_{3\lambda}
-\frac{\mu}3\,\hat{j}_{4\lambda}-\frac{1}2\,\bar{\hat{j}}_{4\lambda}\,\,\,. \label{8-53}
\end{eqnarray}
Expressions \eqref{8-48}, \eqref{8-49} and \eqref{8-50} in \eqref{8-44} give rise an equation for ${c^T}_\lambda$
\begin{eqnarray}
\mu{\bar c^T}_\lambda  +(1+2\mu^2) {c^T}_\lambda =\frac{k}{5\mu}\,\rho_\lambda  \,\,\,, \label{8-54}
\end{eqnarray}
where
\begin{eqnarray}
\rho_\lambda\equiv -\bar l_{1 \lambda}-\bar l_{2 \lambda}+ 2 l_{3 \lambda} -6\mu l_{1 \lambda}-4\mu l_{2 \lambda}-\frac{5}{4}\,\hat{j}_{1\lambda}\,\,\,. \label{8-55}
\end{eqnarray}
Using  \eqref{8-54} and its dual, the second order equation for ${c^T}_\lambda$ can be obtained
\begin{eqnarray}
(\mu^2-\mu_+^2)(\mu^2-\mu_-^2){c^T}_\lambda=(\frac{1+2\mu^2}{20})\rho_\lambda-\frac{1}{20\mu}\bar\rho_\lambda \,\,\,. \label{8-56}
\end{eqnarray}
with a complex poles $\mu_{\pm}^2= -\frac{3}8\pm i\frac{\sqrt{7}}8$, driving to a non-unitary spin 1 propagation. The same story shall be expected for ${w^T}_\lambda$, ${b^T}_\lambda$ and ${v^T}_\lambda$.

\subsection{Spin 0.}

The spin $0$ sector is represented by the action \eqref{8-22}. The field equation for $\psi$ and $c^L$ are $5\psi-2w=c^L=0$, so these fields can be removed from the theory. Then, the two degree of freedom reduced action with interaction is
\begin{eqnarray}
S^{*(0)}[j_1,j_2] =\frac{2}5\,\langle 
-w \Box^{\frac{1}2} \theta +\frac{m}5\,w^2-5m\theta^2-kj_1w-kj_2\theta
\rangle \,\,\,.\label{8-57}
\end{eqnarray}
Particularizing for $j_2=0$ the equation for $w$ is $[\Box+(2m)^2]w=10mkj_1$ and the one-particle exchange amplitude density is
\begin{eqnarray}
\mathcal{A}(w)= 10mk\,j'_1 D(i2m)j_1 \,\,\,,\label{8-58}
\end{eqnarray}
whose residue, $Res(\mathcal{A})>0$ represents an unitary massive tachyon. On the other hand, for $j_1=0$, the field equation for $\theta$ is $[\Box+(2m)^2]w=-\frac{2mk}5j_2$
and one-particle exchange amplitude density is
\begin{eqnarray}
\mathcal{A}(\theta)= -\frac{2mk}5\,j'_2 D(i2m)j_2 \,\,\,,\label{8-59}
\end{eqnarray}
with $Res(\mathcal{A})<0$  means a massive tachyonic ghost.

Focussing our attention on spin $0$, just for an illustration, a way to control these undesirable propagations consists on the construction of counter terms with the help of spin sector operators (appendix B). In this situation one can propose three scalar fields given by 
\begin{eqnarray}
\gamma_a= {{P^{(0)}_a}^\alpha}_{\mu\nu}{\gamma_\alpha}^{\mu\nu}\,\,\,,\label{8-60}
\end{eqnarray}
where ${{P^{(0)}_a}^\alpha}_{\mu\nu}$ are spin $0$ operators. So, the (counter) action with support in Einstein-Weyl space is 
\begin{eqnarray}
S^{(0)}_{counter} =\sum_{a=1}^{3} \alpha_a \langle 
-\frac{1}2\,\partial_\mu \gamma_a \partial^\mu \gamma_a -\frac{m^2_a}2\,\gamma_a\gamma_a
\rangle \,\,\,,\label{8-61}
\end{eqnarray}
for a given real parameters $\alpha_a$ and $m_a$.

Now, the new action would be
\begin{eqnarray}
S'^{(0)}=S^{(0)}+ S^{(0)}_{counter}  \,\,\,,\label{8-63}
\end{eqnarray}
with field equations
\begin{eqnarray}
\alpha_2(\Box-m^2_2)\theta  -\frac{2}5\,\Box^{\frac{1}2}w-4m\theta=0
\,\,\,,\label{8-64}
\end{eqnarray}
\begin{eqnarray}
\alpha_3(\Box-m^2_3)(w+\frac{5}2\,c^L) -\frac{2}5\,\Box^{\frac{1}2}\theta+\frac{4m}5\,w-\frac{8m}5\,\psi=0
\,\,\,,\label{8-65}
\end{eqnarray}
\begin{eqnarray}
\alpha_1(\Box-m^2_1)(\psi+c^L) +4m\psi-\frac{8m}5\,w=0
\,\,\,,\label{8-66}
\end{eqnarray}
\begin{eqnarray}
\alpha_1(\Box-m^2_1)(\psi+c^L) +\frac{5\alpha_3}2\,(\Box-m^2_3)(w+\frac{5}2\,c^L)-4mc^L=0
\,\,\,,\label{8-67}
\end{eqnarray}
and these can be solved in terms of hat-field $\hat\theta \equiv \Box^{-\frac{1}2}\theta$, for example as follows
\begin{eqnarray}
w(\theta)=\frac{5}2\,\big[\alpha_2(\Box-m^2_2)-4m\big]\hat \theta
\,\,\,,\label{8-68}
\end{eqnarray}
\begin{eqnarray}
c^L(\theta)=\big[(-\frac{\alpha_2}4+\frac{1}{4m})\Box+ \frac{m^2_2}4\,\alpha_2+m\big]\hat \theta
\,\,\,,\label{8-69}
\end{eqnarray}
\begin{eqnarray}
\psi(\theta)=\frac{5}{8m}\,\Big[\frac{5\alpha_3}{8m}(1+3m\alpha_1)\Box^2+ (-\frac{15}8\,(m^2_2+m^2_3)\alpha_2\alpha_3- \frac{15}2\,m\alpha_3 + \nonumber \\+\frac{10}3\,m\alpha_2-  \frac{5}{8m}\,m^2_3\alpha_3-\frac{2}5 )\Box -(-\frac{3}4\,m^2_3\alpha_3+\frac{4}5\,m)(\frac{5}2\,m^2_2\alpha_2+10m)    \Big]\hat \theta
\,\,\,,\label{8-70}
\end{eqnarray}
and
\begin{eqnarray}
\Big[A\Box^3+B\Box^2 +C\Box+D\Big]\hat \theta=0
\,\,\,,\label{8-71}
\end{eqnarray}
with parameters defined by
\begin{eqnarray}
A\equiv \frac{25\alpha_1\alpha_3}{64m}\,\big(\frac{1}{m}+3\alpha_1\big)
\,\,\,,\label{8-72}
\end{eqnarray}
\begin{eqnarray}
B\equiv\frac{5\alpha_1}{8m}\,\big[-\frac{15}8\,(m^2_2+m^2_3)\alpha_2\alpha_3- \frac{15}2\,m\alpha_3 +\frac{10}3\,m\alpha_2-  \frac{5}{8m}\,m^2_3\alpha_3+\nonumber \\
 +\alpha_3(-m^2_1\alpha_1+4m)(\frac{5}{8m\alpha_1}+2) -\frac{2m\alpha_2}5   \big]
\,\,\,,\label{8-73}
\end{eqnarray}
\begin{eqnarray}
C\equiv\frac{5\alpha_1}{8m}\,\Big[(\frac{3}4\,m^2_3\alpha_3+\frac{4m}5)(\frac{5}2\,m^2_2\alpha_2+10m)+(m^2_1-\frac{4m}{\alpha_1})\big[\frac{15}8\,(m^2_2+m^2_3)\alpha_2\alpha_3+\nonumber \\     + \frac{15}2\,m\alpha_3 -\frac{10}3\,m\alpha_2+  \frac{5}{8m}\,m^2_3\alpha_3
 +\frac{2}5 \big] +\frac{4m}{25}\,(\frac{5}2\,m^2_2\alpha_2+10m)+\nonumber \\ +\frac{4m\alpha_2}{\alpha_1}\,(\frac{m^2_1}{10}\,\alpha_1+\frac{8m}5) -\frac{2m^2_1}5\Big]
\,\,\,,\nonumber \\  \label{8-74}
\end{eqnarray}
\begin{eqnarray}
D\equiv\frac{5}{8m}\,(\frac{5}2\,m^2_2\alpha_2+10m)\Big[(m^2_1\alpha_1-4m)(-\frac{3}4\,m^2_3\alpha_3+\frac{4m}5)+\nonumber \\ +\frac{m^2_1\alpha_1}{10}-\frac{8m}5\Big]
\,\,\,. \label{8-75}
\end{eqnarray}
In order to reach a causal propagation and to avoid an undesirable unphysical particle spectre we demand the following conditions on parameters
\begin{eqnarray}
A=B=0\,\,\,,\,\,\,\,\frac{D}C<0
\,\,\,. \label{8-76}
\end{eqnarray}
Next, one can to identify three classes of counter-actions  \eqref{8-61} at least: $\alpha_1=0$,  $\alpha_3=0$ and $\alpha_1=-\frac{1}{3m}$.
Just for an illustration, the first one, this means $\alpha_1=0$ induces    
\begin{eqnarray}
\alpha_3=0
\,\,\,, \label{8-77}
\end{eqnarray}
\begin{eqnarray}
C=\frac{13}3\,m\alpha_2-1
\,\,\,, \label{8-78}
\end{eqnarray}
\begin{eqnarray}
D=-mm^2_2\alpha_2-(2m)^2
\,\,\,, \label{8-79}
\end{eqnarray}
and the condition for causal propagation, $ D/C<0$ means
\begin{eqnarray}
(2m)^2\Big[\frac{1+\frac{m^2_2\alpha_2}{4m}}{1-\frac{13m\alpha_2}3}\Big]<0
\,\,\,, \label{8-80}
\end{eqnarray}
characterizing the $(m_2,\alpha_2)$-family of possible counter-actions. 

So, there is no an unique way to control ill spin $0$ propagations.  Just as there are ways to eliminate non-physical spin $0$, there are also ways to control non-physical spins $1$ and $2$ particles. Added to this is the incorporation of first and second order terms related to spin $1$ and $2$ and even higher order actions in the Riemann curvature as discussed in section 3. A sufficiently general action candidate for a modified massive gravity in Einstein-Weyl is
\begin{eqnarray}
S'_{TMGEW} = S_{TMGEW} + S^{(0)}_{counter}+ S^{(1)}_{counter}+S^{(2)}_{counter}+\langle F^{(2)}(R)\rangle \,\,\,, \label{8-81}
\end{eqnarray}
where $F^{(2)}(R)$ is a suitable and quadratical in curvature lagrangian density built in consistence with the existence of conformally flatness solutions and $S_{TMGEW}$ is defined at \eqref{8-7}. In \eqref{8-81}, the (counter) actions for spin $1$ and $2$ sectors could be the following selfdual models
\begin{eqnarray}
S^{(1)}_{counter}=\sum_{a=1}^{3} \beta_a\langle  \varepsilon^{\mu\nu\lambda}\gamma_{a\mu} D_\nu \gamma_{a\lambda}-m^{(1)}_a\gamma_{a\mu} {\gamma_a}^\mu \rangle \,\,\,,\label{8-82}
\end{eqnarray}
\begin{eqnarray}
S^{(2)}_{counter}=\sum_{a=1}^{2} \sigma_a\langle  \varepsilon^{\mu\nu\lambda}\gamma_{a\mu\beta}D_\nu{\gamma_{a\lambda}}^\beta- m^{(2)}_a (\gamma_{a\mu\nu}{\gamma_a}^{\nu\mu}-{{\gamma_a}^\mu}_\mu{{\gamma_a}^\nu}_\nu)\rangle \,\,\,,\label{8-83}
\end{eqnarray}
where parameters are adequately defined and
\begin{eqnarray}
\gamma_{a\nu}= {{P^{(1)}_a}^\alpha}_{\mu\beta\nu}{\gamma_\alpha}^{\mu\beta}\,\,\,,\,\,\,\,\,\,\,\,\,\,a=1,2,3\,\,\,\,\,\,\,,\label{8-84}
\end{eqnarray}
\begin{eqnarray}
\gamma_{a\nu\rho}= {{P^{(2)}_a}^\alpha}_{\mu\beta\nu\rho}{\gamma_\alpha}^{\mu\beta}\,\,\,,\,\,\,\,\,\,\,\,\,\,a=1,2\,\,\,\,\,\,\,,\label{8-85}
\end{eqnarray}
with ${{P^{(1)}_a}^\alpha}_{\mu\beta\nu}$ and ${{P^{(2)}_a}^\alpha}_{\mu\beta\nu\rho}$ as the spin sector operators for spin $1$ and $2$, respectively (appendix B). Obviously the terms $S^{(0)}_{counter}+ S^{(1)}_{counter}+S^{(2)}_{counter}$ are only supported in Einstein-Weyl Space and they do not affect the degree of freedom counting in a Riemann-Cartan Space.

\section{Conclusion.}

Certain classical properties of the $F(R)$ theories have been briefly reviewed and explored, within which it is well known that those of the quadratic type have been presented as good candidates to correct certain disadvantages regarding the presence of non-physical propagations in gravity theories.  The relevance of elaborating the discussion on the reduction of the general geometry of the MAG is important at the same time that Einstein's gravitation could be considered as test evidence for any modified gravity model at some limit. Hence, the existence or not of conformally flat solutions must be an elementary requirement of any generalized model that is proposed.

In the case of topologically massive gravity, we have tried to stay as faithful as possible to DJT's original idea, keeping the two original pieces, such as the Hilbert-Einstein Lagrangian density and the Chern-Simons density, and extending their theory to the level of non-pseudo-Riemannian geometry  without introducing new Lagrangian terms. Through the Riemann-Cartan reduction path ($T\neq 0$, $Q=0$) we verify the same degrees of freedom as the Mielke-Baekler model but from a point of view of a modified selfdual model, while in the  Einstein-Weyl scenario ($T=0$, $Q\neq 0$) the abundance of particles brought multiple non-unitary propagations which warrants an evaluation of the model itself. Obviously this has to bring stability problems in any classical non-trivial vacuum solution that is tested. The advantage is that there is apparently no unique way to correct the above. In fact, actions of the quadratic type in curvature are candidates to control non-unitary propagations as in the case of TMGEW. A classical criterion was envisioned to choose the possible families of theories $F^{(2)}(R)$ that could help to control non-physical propagations based on requirement that conformally planar solutions must be included. A criterion that could be extended beyond $2+1$ dimensions in other models. We have also verified that not only higher order curvature theories works, but they are complemented by other possible selfdual type of spin $1$ and $2$.

Several problems arise as the question about whether the modified selfdual model found is unique or not, the construction of a topologically massive gravity model that overcomes the problems shown in Einstein-Weyl, the study of classical and quantum stability, among others will be addressed elsewhere.


\section*{Acknowledgments}

One of the authors (R. G.) dedicates this work to the memory of Mrs. Gatu for her unconditional support and affection.

\appendix

\section{Diffeomorphisms.}

A transformation under difeomorphisms is a smooth deformation (continuous, differentiable and invertible) of a manifold that is manifested by the continuous and differentiable change of the coordinates that label the points of space-time. Thus, we will think about transformations of the form ${x'}^\mu=x^\mu - \xi^\mu(x)$, where $\xi^\mu(x)$ are infinitesimal functions ($0<|\xi^\mu(x)|<<1$) continuous and derivable. Let's see how
infinitesimal diffeomorphism on coordinates induces functional transformation on fields. Under the group of general transformations a scalar field transforms like ${\phi'}(x')=\phi(x)$, then a diffeomorphism variation means 
\begin{align}
\delta_\xi\phi(x)&\equiv{\phi'}(x)-\phi(x) \nonumber \\
&={\phi'}(x'-\xi)-\phi(x)\nonumber \\
&={\phi'}(x')+\frac{1}{1!}\partial'_\mu{\phi'}(x')\xi^\mu+\mathcal{O}(\xi^2)-\phi(x)\nonumber \\
&={\phi'}(x')-\phi(x)+\frac{1}{1!}\big(\partial_\mu{\phi'}(x)-\frac{1}{1!}\partial_\alpha \partial_\mu{\phi'}(x)\xi^\alpha+...\big)\xi^\mu(x)+...\nonumber \\
&={\phi'}(x')-\phi(x)+\frac{1}{1!}\partial_\mu\big(\phi(x)+\delta_\xi\phi(x)\big)\xi^\mu(x)+...\nonumber \\
&=\xi^\mu (x)\partial_\mu\phi(x)\,\,.\,\,\label{A4}
\end{align}

A rank $1$ tensor transforms as ${t'}^\mu(x')={U^\mu}_\alpha t^\alpha$ where $U\in GL(N,R)$. For an infinitesimal transformation under diffeomorphisms the elements of $GL(N,R)$ are ${U^\mu}_\alpha\mid_{\xi}={\delta^\mu}_\alpha- \partial_\alpha\xi^\mu $ and induce the variation of diffeomorphisms on field $t^\mu$ as follows
\begin{align}
\delta_\xi t^\mu(x) =\xi^\alpha (x)\partial_\alpha t^\mu(x)- \partial_\alpha\xi^\mu (x) t^\alpha (x) \,\,.\,\,\label{A5}
\end{align}

In the same way, the following transformation rules can be shown 

(i) a rank $1$ (contravariant) tensor
\begin{align}
\delta_\xi t_\mu(x) =\xi^\alpha (x)\partial_\alpha t_\mu(x)+ \partial_\mu\xi^\alpha (x) t_\alpha (x) \,\,. \,\,\label{A6}
\end{align}

(ii) a rank $2$  tensor
\begin{align}
\delta_\xi t_{\mu\nu}(x) =\xi^\alpha (x)\partial_\alpha t_{\mu\nu}(x)+ \partial_\mu\xi^\alpha (x) t_{\alpha\nu} (x)+ \partial_\nu\xi^\alpha (x) t_{\mu\alpha} (x) \,\,.\,\,\label{A7}
\end{align}

(iii) a scalar density of ''weight'' $W$ (i. e.,  $\psi' (x')= det {\mid U\mid}^W \psi (x)$)
\begin{align}
\delta_\xi \psi (x)= \xi^\mu (x)\partial_\mu\psi(x) -W\partial_\mu\xi^\mu (x)\psi(x)  \,\,. \,\,\label{A8}
\end{align}

(iv) Levi-Civita density in $2+1$ dimension 
\begin{align}
\delta_\xi \epsilon^{\mu\nu\lambda} =\partial_\alpha\xi^\alpha  \epsilon^{\mu\nu\lambda}-\partial_\alpha\xi^\mu \epsilon^{\alpha\nu\lambda} -\partial_\alpha\xi^\nu \epsilon^{\alpha\lambda\mu}-\partial_\alpha\xi^\lambda \epsilon^{\alpha\mu\nu} \,\,. \,\,\label{A9}
\end{align}

\section{TLt decomposition. Spin sector operators.}

The covariant Transverse-Longitudinal-traceless decomposition (TLt) for some tensor objects are as follows. For a range tensor $ 1 $ we write
\begin{align}
V_\mu = {V^T}_\mu+ \hat{\partial }_\mu V^L\,\,, \,\,\label{B1}
\end{align}
where $\partial^\mu {V^T}_\mu=0$ and $\hat{\partial }^\mu\equiv \frac{\partial ^\mu}{\Box{}
^{\frac{1}{2}} }$.

For a symmetric rank $2$ tensor (i. e., $h_{\mu \nu }= h_{\nu \mu }$) we get
\begin{equation}
h_{\mu \nu }\equiv {h^{Tt}}_{\mu \nu }+\hat\partial _\mu
{a^T}_\nu +\hat\partial _\nu {a^T}_\mu + \hat\partial _\mu\hat\partial
_\nu\phi +\eta _{\mu \nu} \,\psi \,\,, \,\,\label{B2}
\end{equation}
with the supplementary conditions
\begin{equation}
{{h^{Tt}}_\mu}^\mu =0 \,\,, \,\,\label{B3}
\end{equation}
\begin{equation}
\hat\partial^\mu{h^{Tt}}_{\mu \nu }=0
\,\,, \,\,\label{B4}
\end{equation}
\begin{equation}
\hat\partial _\mu {a^T}^\mu=0
\,\,. \,\,\label{B5}
\end{equation}
Objects $\phi$ and $\psi$ are scalar fields.

For a rank $3$ tensor semisymmetric and traceless in the two last indexes, this means $\omega_{\alpha\mu\nu}=\omega_{\alpha\nu\mu}$ and ${{\omega_\alpha}^\mu}_\mu=0$ we get
\begin{eqnarray}
\omega_{\alpha\mu\nu}={S^{Tt}}_{\alpha\mu\nu}+\hat\partial_{(\alpha}{s^{Tt}}_{\mu\nu)}+2\hat\partial_{\{\mu}{\tau^{Tt}}_{\alpha\nu\}}
+[\hat\partial_{(\alpha}\hat\partial_\mu -\frac{\eta_{(\alpha\mu}}5]{w^T}_{\nu)}+2\hat\partial_{\{\mu}\hat\partial_\alpha{b^T}_{\nu\}}+
\nonumber \\  +2 {\epsilon_{\alpha\{\mu}}^\lambda(\hat\partial_{\nu\}}{v^T}_\lambda+\hat\partial_\lambda{v^T}_{\nu\}})+
[\hat\partial_\alpha\hat\partial_\mu\hat\partial_\nu -\frac{\eta_{(\alpha\mu}}5\,\hat\partial_{\nu)}]w+2 {\epsilon_{\alpha\{\mu}}^\lambda\hat\partial_{\nu\}}\hat\partial_\lambda\theta+\nonumber \\ -2(\hat\partial_\alpha\hat\partial_\mu\hat\partial_\nu -\eta_{\alpha\{\mu}\hat\partial_{\nu\}})\psi
 \,\,\,,   \nonumber \\
\label{B6}
\end{eqnarray}
where ${S^{Tt}}_{\alpha\mu\nu}$ is a totally symmetric Tt tensor, objects ${s^{Tt}}_{\mu\nu}$ and ${\tau^{Tt}}_{\alpha\nu}$ are symmetric Tt tensors, etc.
All these TLt components can be solved using the {\it spin sector operators}. Let us introduce them as follows

\vskip .1truein

(i) Spin $0$

\begin{eqnarray}
{{P^{(0)}_1}^\alpha}_{\mu\beta}=\frac{1}{2}\,{\delta^\alpha}_{\{\mu}\hat\partial_{\beta\}}\,\,\,,\,\,\label{B7}
\end{eqnarray}
\begin{eqnarray}
{{P^{(0)}_2}^\alpha}_{\mu\beta}=\frac{1}{2}\,{\Theta^\alpha}_{\{\mu}\hat\partial_{\beta\}}\,\,\,,\,\,\label{B8}
\end{eqnarray}
\begin{eqnarray}
{{P^{(0)}_3}^\alpha}_{\mu\beta}=\frac{5}{2}\,\hat\partial^\alpha\hat\partial_\mu\hat\partial_\beta\,\,\,,\,\,\label{B9}
\end{eqnarray}

(ii) Spin $1$

\begin{eqnarray}
{{P^{(1)}_1}^\alpha}_{\mu\beta\nu}={\delta^\alpha}_{\{\mu}P_{\nu\beta\}}\,\,\,,\,\,\label{B10}
\end{eqnarray}
\begin{eqnarray}
{{P^{(1)}_2}^\alpha}_{\mu\beta\nu}=\frac{1}{3}\,{\Theta^\alpha}_{\{\mu}P_{\nu\beta\}}\,\,\,,\,\,\label{B11}
\end{eqnarray}
\begin{eqnarray}
{{P^{(1)}_3}^\alpha}_{\mu\beta\nu}=-\frac{5}6\,{{P^{(1)}_1}^\alpha}_{\mu\beta\nu}-\frac{1}2\,{{P^{(0)}_3}^\alpha}_{\mu\beta}\hat\partial_\nu+
\frac{5}{12}\,{\eta_\nu}^{(\alpha}\hat\partial_\mu\hat\partial_{\beta)}\,\,\,,\,\,\label{B12}
\end{eqnarray}

(iii) Spin $2$

\begin{eqnarray}
{{P^{(2)}_1}^\alpha}_{\mu\beta\rho\nu}=\frac{1}5\,\big[\eta_{\rho\nu}+\hat\partial_\rho\hat\partial_\nu\big]{{P^{(0)}_3}^\alpha}_{\mu\beta}
+(\eta_{\{\mu\rho}\eta_{\beta\}\nu}-2\eta_{<\mu\{\rho}\hat\partial_{\nu\}}\hat\partial_{\beta>})\hat\partial^\alpha\,\,\,,\,\,\label{B13}
\end{eqnarray}
\begin{eqnarray}
{{P^{(2)}_2}^\alpha}_{\mu\beta\rho\nu}={\delta^\alpha}_{\{\rho}P_{<\mu\nu\}}\hat\partial_{\beta>}-\eta_{\{\mu<\rho}P_{\beta\}\nu>}
-P_{\rho\nu}{{P^{(0)}_1}^\alpha}_{\mu\beta}\,\,\,,\,\,\label{B14}
\end{eqnarray}

(iv) Spin $3$

\begin{eqnarray}
{{P^{(3)}}^\alpha}_{\mu\beta\lambda\rho\nu}=\frac{1}3\,P_{(\lambda\rho}\hat\partial_{\nu)}{{P^{(0)}_1}^\alpha}_{\mu\beta}
-\frac{1}{10}\,\big[\eta_{(\lambda\rho}\hat\partial_{\nu)}+\hat\partial_\lambda\hat\partial_\rho\hat\partial_\nu\big]{{P^{(0)}_3}^\alpha}_{\mu\beta}
+\nonumber \\ +\frac{1}3\,\eta_{(\rho\{\mu}\eta_{\nu\beta\}}{P^\alpha}_{\lambda)}-\frac{1}6\,{\delta^\alpha}_{\{\mu}\eta_{\beta\}(\nu}P_{\lambda\rho)}
-\frac{1}3\,\big[{\delta^\alpha}_{(\lambda}\eta_{\{\beta\nu}+{\delta^\alpha}_{(\nu}\eta_{\{\beta\lambda}\big]\hat\partial_{\rho)}\hat\partial_{\mu\}}+\nonumber \\ + \big[\frac{\eta_{(\rho\lambda}}3+\hat\partial_{(\rho}\hat\partial_\lambda\big]
\big[ \frac{{\delta^\alpha}_{\nu)}}4\,\hat\partial_{\{\beta}+\frac{\eta_{\nu)\{\beta}}2\,\hat\partial^\alpha\big]\hat\partial_{\mu\}}\,\,\,,\,\,\,\,\,\,\label{B15}
\end{eqnarray}
where ${\Theta_\mu}^\nu\equiv {\epsilon_\mu}^{\sigma\nu}\hat\partial_\sigma$ is a parity operator,  ${P_\mu}^\lambda= {\delta_\mu}^\lambda-\hat\partial_\mu\hat\partial^\lambda$ is a transverse projector, index symbols $\{.\}$ or $<.>$ means independent symmetrizations and $(.)$ represents the cyclic sum over three indexes. For example, 
\begin{eqnarray}
{S^{Tt}}_{\lambda\rho\nu}= {{P^{(3)}}^\alpha}_{\mu\beta\lambda\rho\nu} \,\,{\omega_\alpha}^{\mu\beta}
 \,\,\,,\,\,  
\label{B16}
\end{eqnarray}
etc.

\end{document}